\documentclass[longauth,traditabstract]{aa} 
\usepackage{graphicx}
\usepackage{txfonts}
\usepackage{natbib}
\newcommand{\oii}{\mbox{[O\,{\scriptsize II}]}}

\begin{document}

   \title{The VIMOS Public Extragalactic Redshift Survey (VIPERS):}
   \subtitle{A quiescent formation of massive red-sequence galaxies over the past 9 Gyr
   \thanks{Based on observations collected at the European Southern Observatory,
    Cerro Paranal, Chile, using the Very Large Telescope under programs
    182.A-0886 and partly 070.A-9007. Also based on observations obtained with
    MegaPrime/MegaCam, a joint project of CFHT and CEA/DAPNIA, at the
    Canada-France-Hawaii Telescope (CFHT), which is operated by the National
    Research Council (NRC) of Canada, the Institut National des Sciences de
    l'Univers of the Centre National de la Recherche Scientifique (CNRS)
    of France, and the University of Hawaii. This work is based in part on data
    products produced at TERAPIX and the Canadian Astronomy Data Centre as part
    of the Canada-France-Hawaii Telescope Legacy Survey, a collaborative project
    of NRC and CNRS. The VIPERS web site is http://www.vipers.inaf.it/.}}

\titlerunning{VIPERS - A quiescent formation of massive red-sequence galaxies since $z\sim 1.3$}

\author{
A.~Fritz\inst{1}
\and M.~Scodeggio\inst{1}
\and O.~Ilbert\inst{2}
\and M.~Bolzonella\inst{3}           
\and I.~Davidzon\inst{3,4}
\and J.~Coupon\inst{5}
\and B.~Garilli\inst{1,2}     
\and L.~Guzzo\inst{6,7}
\and G.~Zamorani\inst{3}
\and U.~Abbas\inst{8}
\and C.~Adami\inst{2}
\and S.~Arnouts\inst{9,2}
\and J.~Bel\inst{10}
\and D.~Bottini\inst{1}
\and E.~Branchini\inst{11,12,13}
\and A.~Cappi\inst{3,14}
\and O.~Cucciati\inst{3}           
\and G.~De Lucia\inst{15}
\and S.~de la Torre\inst{16}
\and P.~Franzetti\inst{1}
\and M.~Fumana\inst{1}
\and B.~R.~Granett\inst{6}
\and A.~Iovino\inst{6}
\and J.~Krywult\inst{17}
\and V.~Le Brun\inst{2}
\and O.~Le F\`evre\inst{2}
\and D.~Maccagni\inst{1}
\and K.~Ma{\l}ek\inst{18}
\and F.~Marulli\inst{4,19,3}
\and H.~J.~McCracken\inst{20}
\and L.~Paioro\inst{1}
\and M.~Polletta\inst{1}
\and A.~Pollo\inst{21,22}
\and H.~Schlagenhaufer\inst{23,24}
\and L.~A.~M.~Tasca\inst{2}
\and R.~Tojeiro\inst{25}
\and D.~Vergani\inst{26}
\and A.~Zanichelli\inst{27}
\and A.~Burden\inst{25}
\and C.~Di Porto\inst{3}
\and A.~Marchetti\inst{28,6} 
\and C.~Marinoni\inst{10}
\and Y.~Mellier\inst{20}
\and L.~Moscardini\inst{4,19,3}
\and R.~C.~Nichol\inst{25}
\and J.~A.~Peacock\inst{16}
\and W.~J.~Percival\inst{25}
\and S.~Phleps\inst{23}
\and M.~Wolk\inst{20}
}

\authorrunning{A.~Fritz et al.}

\offprints{Alexander Fritz, VIPERS fellow \\ \email{afritz@iasf-milano.inaf.it}}

\institute{
 INAF - Istituto di Astrofisica Spaziale e Fisica Cosmica (IASF) Milano, via E. Bassini 15, 20133 Milano, Italy
\and Aix Marseille Universit\'e, CNRS, LAM (Laboratoire d'Astrophysique de Marseille) UMR 7326, 13388, Marseille, France  
\and INAF - Osservatorio Astronomico di Bologna, via Ranzani 1, I-40127, Bologna, Italy 
\and Dipartimento di Fisica e Astronomia - Universit\`{a} di Bologna, viale Berti Pichat 6/2, I-40127 Bologna, Italy 
\and Institute of Astronomy and Astrophysics, Academia Sinica, P.O. Box 23-141, Taipei 10617, Taiwan
\and INAF - Osservatorio Astronomico di Brera, Via Brera 28, 20122 Milano, via E. Bianchi 46, 23807 Merate, Italy 
\and Dipartimento di Fisica, Universit\`a di Milano-Bicocca, P.zza della Scienza 3, I-20126 Milano, Italy 
\and INAF - Osservatorio Astrofisico di Torino, 10025 Pino Torinese, Italy 
\and Canada-France-Hawaii Telescope, 65--1238 Mamalahoa Highway, Kamuela, HI 96743, USA 
\and Aix-Marseille Universit\'e, CNRS, CPT (Centre de Physique  Th\'eorique) UMR 7332, F-13288 Marseille, France 
\and Dipartimento di Matematica e Fisica, Universit\`{a} degli Studi Roma Tre, via della Vasca Navale 84, 00146 Roma, Italy 
\and INFN, Sezione di Roma Tre, via della Vasca Navale 84, I-00146 Roma, Italy 
\and INAF - Osservatorio Astronomico di Roma, via Frascati 33, I-00040 Monte Porzio Catone (RM), Italy 
\and Laboratoire Lagrange, UMR7293, Universit\'e de Nice Sophia-Antipolis,  CNRS, Observatoire de la C\^ote d'Azur, 06300 Nice, France 
\and INAF - Osservatorio Astronomico di Trieste, via G. B. Tiepolo 11, 34143 Trieste, Italy 
\and SUPA, Institute for Astronomy, University of Edinburgh, Royal Observatory, Blackford Hill, Edinburgh EH9 3HJ, UK 
\and Institute of Physics, Jan Kochanowski University, ul. Swietokrzyska 15, 25-406 Kielce, Poland 
\and Department of Particle and Astrophysical Science, Nagoya University, Furo-cho, Chikusa-ku, 464-8602 Nagoya, Japan 
\and INFN, Sezione di Bologna, viale Berti Pichat 6/2, I-40127 Bologna, Italy 
\and Institute d'Astrophysique de Paris, UMR7095 CNRS, Universit\'{e} Pierre et Marie Curie, 98 bis Boulevard Arago, 75014 Paris, France 
\and Astronomical Observatory of the Jagiellonian University, Orla 171, 30-001 Cracow, Poland 
\and National Centre for Nuclear Research, ul. Hoza 69, 00-681 Warszawa, Poland 
\and Max-Planck-Institut f\"{u}r Extraterrestrische Physik, D-84571, Garching b. M\"{u}nchen, Germany 
\and Universit\"{a}tssternwarte M\"{u}nchen, Ludwig-Maximillians Universit\"{a}t, Scheinerstr. 1, D-81679 M\"{u}nchen, Germany 
\and Institute of Cosmology and Gravitation, Dennis Sciama Building, University of Portsmouth, Burnaby Road, Portsmouth, PO1 3FX 
\and INAF - Istituto di Astrofisica Spaziale e Fisica Cosmica Bologna, via Gobetti 101, I-40129 Bologna, Italy 
\and INAF - Istituto di Radioastronomia, via Gobetti 101, I-40129, Bologna, Italy 
\and Universit\`{a} degli Studi di Milano, via G. Celoria 16, 20130 Milano, Italy 
}

\date{Received -; accepted -}
 
  \abstract
  {We explore the evolution of the Colour-Magnitude Relation (CMR) and Luminosity
Function (LF) at $0.4$$<$$z$$<$$1.3$ from the VIMOS Public Extragalactic
Redshift Survey (VIPERS) using $\sim$45,000 galaxies with precise spectroscopic
redshifts down to $i'_{AB}$$<$$22.5$ over $\sim$10.32~deg$^{2}$ in two fields.
 From $z=0.5$ to $z=1.3$ the LF and CMR are well defined for different galaxy
populations  and $M^{*}_{B}$ evolves by $\sim$$1.04(1.09)\pm0.06(0.10)$~mag 
for the total (red) galaxy sample. We compare different criteria for 
selecting early-type galaxies: (1) a fixed cut in rest-frame 
$(U-V)$ colours, (2) an evolving cut in $(U-V)$ colours, (3) a rest-frame 
$(NUV-r')-(r'-K)$ colour selection, and (4) a spectral-energy-distribution 
classification. The completeness and contamination varies for the different 
methods and with redshift, but regardless of the method we measure a 
consistent evolution of the red-sequence (RS). 
Between $0.4<z<1.3$ we find a moderate evolution of the RS intercept of
$\Delta(U-V)=0.28\pm0.14$~mag, favouring exponentially declining 
star formation (SF) histories with SF truncation at $1.7\leq z\leq2.3$.
Together with the rise in the number density of red galaxies by 0.64~dex
since $z=1$, this suggests a rapid build-up of massive galaxies 
($M_{\star}>10^{11}M_{\odot})$ and 
expeditious RS formation over a short period of $\sim$1.5~Gyr starting before
$z=1$. This is supported by the detection of ongoing SF in early-type galaxies 
at $0.9<z<1.0$, in contrast with the quiescent red stellar populations 
of early-type galaxies at $0.5<z<0.6$. There is an increase in the observed CMR 
scatter with redshift, which is two times larger than observed in galaxy 
clusters and at variance with theoretical model predictions. We discuss 
possible physical mechanisms that support the observed evolution of the red 
galaxy population. Our findings point out that massive galaxies have 
experienced a sharp SF quenching at $z\sim1$ with only limited additional 
merging. In contrast, less-massive galaxies experience a mix of SF truncation
and minor mergers which build-up the low- and intermediate-mass end of the
CMR.}


   \maketitle


\section{Introduction}\label{intro} 

Early-type galaxies are a unique class with rather simple and homogeneous
global properties, like morphology, structure, colours, kinematics, and 
stellar population content. Observationally it is well known that tight correlations
exist among these properties, which are often called fundamental relations 
\citep{B59,FJ76,Dre80,DD87,Dre87,VS77}\footnote{Although the actual 
picture is somewhat more complicated because of the observed differences 
between elliptical and lenticular (S0) galaxies, here we treat them all as a 
single galaxy class.}. However, both the formation processes and the subsequent
evolution of these systems with redshift are still uncertain and actively debated.

The most commonly accepted evolutionary scenario for all types of galaxies
since $z=1$ is the so-called downsizing scenario \citep{GPB96,CSHC96}, with
massive galaxies forming the bulk of their stars within short, highly-peaked
star formation periods at earlier epochs, whereas less-massive galaxies have
delayed star formation histories which are extended over a longer time
period \citep{GPB96,TMBO05,NSHW05,Jim07,FDLMSS09}. Independent 
evidence supports this scenario. The tight Fundamental Plane relations which
exist in both cluster and field environments suggest a higher/lower formation
redshift of the stellar content in massive-/less-massive E/S0s at 
$z_{\rm f}>2/z_{\rm f}\lesssim1$ \citep{FZBSD05,FBZ09,FJSC09,SVCL05,Tre05,vdW08}.
The specific star formation rates (sSFRs) are high/low for low/high-mass
galaxies at $0$$<$$z$$<$$2$, but inverse trends are found at $z$$>$$2$ 
\citep{Feu05,Jun05}. Furthermore, luminous E/S0s show higher mass-to-light 
ratios and different initial mass function (IMF) than their low-massive 
counterparts \citep[e.g.,][]{Fon04,Cap06,Cap12} and there is a rapid decrease
of massive post-starburst galaxies with cosmic time \citep{GDDSLB06,Ver08}.

Such a downsizing scenario for galaxy evolution has some difficulties to be 
included within the hierarchical structure formation framework implied by 
the standard $\Lambda$CDM cosmological model. Simulations show that in 
this framework galaxies assemble their mass continously through mergers of
sub-units over cosmic time, with a mass-dependent evolution of massive 
E/S0s which form more than half of their mass at very late epochs of $z<1$. 
Specifically, semi-analytic models based on the hierarchical
merger trees of dark-matter halos fail in matching the history of
formation and abundance of massive red galaxies, unless a specific feedback
mechanism is included \citep[e.g.,][]{DLSWC06,DLJB07,DLB12}.  Even with these
ingredients, models have difficulty in reproducing the observed weak evolution
since $z\sim 1$ of the bright and massive ends ($M_{\star}>10^{11}M_{\odot}$)
of the early-type galaxy luminosity and stellar mass functions (after
correction for passive evolution), in contrast to the faster evolution of 
less-massive systems \citep{Bel04b,Bor06,Bun06,Cim2006,Fab07}.
Similarly, the number density of luminous and massive early-type galaxies 
remains relatively constant over the past $\sim$8 Gyr ($z$$\sim$$0.8$),
whereas less-luminous (low-mass) systems show a growth over
the same time period \citep[e.g.,][]{Bun05,Cim2006,Bun06,Con07,Sca07,Cas11,IMCLF13}.

Since the study of fundamental relations can be very demanding in terms of 
observations and of telescope time, large galaxy surveys have often adopted 
galaxy colour as the primary parameter to use in the study of galaxy evolution 
via more economical photometric relations, like the colour-magnitude or the 
colour-stellar mass relation. It was thus discovered that galaxies exhibit a 
segregation in luminosity and mass between red, passive early-type (E/S0) 
galaxies, and blue, star forming late-type ones 
\citep{DG76,SBT85,Kau03,Kau04,B2dF04,Bel04b}, and also a strong bimodal
distribution in their properties, like colour, size, star formation, 
luminosity/mass function 
\citep[e.g.,][]{Stra01,Im02,MH02,Hog02,Bel03,Bla03LF,Fon04,B2dF04,Bel04b,Kau04,Wei05,WFK06,Bro07,WNE062DF,Fab07,CEF08,PBZ10,CEK12}.
However, the origin and nature of the observed bimodality and the downsizing 
effect in galaxy properties represents a challenge to the models. Possible 
galaxy evolution models that could provide some explanation for these bimodal
distributions include the self-regulation of star formation processes from 
supernovae feedback (particularly effective in dark matter halo masses below 
$M_{h}\sim5\times10^{11}M_{\odot}$), virial shock heating 
\citep{DB2006,Cat06}, and/or star formation quenching due to Active Galactic
Nuclei (AGN) feedback \citep{Gra04,Men05,Men06,DB2006,Sch06}.   

The bimodal distribution in optical colours of galaxies is mainly a 
consequence of the bulk of early-type galaxies forming a tight 
sequence within the colour-magnitude space, originally termed as 
the ``red-envelope'' \citep{VS77,OCon88,Ell88},
but now known as the  ``red-sequence'' \citep[RS,][]{GLCY98}. The 
RS has been used as a marker in the search of clusters of galaxies 
in the nearby \citep[e.g.,][BLE92]{BLE92a,BLE92b,GBM96,Sco01,LCBY04}
and in the distant universe, up to $z$$\sim$$1$, using optical multi-band
photometry 
\citep{AES91,AEC93,SED95,SED98,RS95,GBM96,ESD97,Bow98,vDFK98,KAB98,GLCY98,vDFK00,Bla03,FZBSD05,Tan05,YHL05,CGF07}.
Recently, the combination of optical, near-infrared (NIR) and/or mid-infrared
(MIR) {\em Spitzer} photometry allowed extending the RS technique 
to detect high-redshift clusters at $1.2\lesssim z\lesssim2.2$ 
\citep{Wil09,SRP10,Dem10,AH11}. The detailed properties of the RS, instead, 
have been used to study the formation and evolution of massive quies\-cent 
galaxies since redshift of $z\sim1$ \citep{Bel04b,Fra07,ECDFSRS,Tan05,Wei05}.
The RS has been demonstrated to exist up to at least $z\sim1.5$ 
\citep{Fra07,W09UVJ,Nic11} and there are suggestions that it might be already
in place at redshift $z\sim2$, but these studies are based on multi-band 
photometry only \citep{Gia05,Tay09,Whi10}.
Still, the most common use of the RS is in the separation of quiescent, 
predominantly red-coloured galaxies from the bulk of the star-forming,  
predominantly blue-coloured galaxies, and from galaxies in the transition zone 
between the blue cloud and the RS (so-called ``green valley''). Unfortunately, 
the operational definition of the RS is not uniform in the literature.
Since galaxies evolve, and the ancestors of the present-day early-type 
galaxies were different (types of) galaxies at high redshift 
\citep{DF01} which underwent a number of transformations 
as they evolved to the final properties they have today, the definition of
a selection criterion for the study of the RS is a complex task, and different
RS definitions can result in rather different galaxy samples, making the 
comparison among various analysis a challenging task.

Another complication is that the volumes covered by deep redshift surveys 
have been so far too small to guarantee adequate sampling of the very rare 
objects on the bright (massive) end of the luminosity (mass) function. 
Therefore, the properties and the contribution of the global galaxy population
to the RS is highly uncertain. Consequently, despite the progress described 
above, there is still significant uncertainty as to how and when the global RS
of the overall population of galaxies integrated over all environments has 
emerged. The data used in this work represent a major step forward in this
direction, being based on nearly 50,000 galaxies from the Public Data
Release 1 (PDR-1) of the VIMOS Public Extragalactic Redshift Survey  
\citep[VIPERS,][]{GSG13}.  The VIPERS data are used to measure the
evolution of the luminosity function (LF) and the colour-magnitude
relation (CMR) of the galaxy population over the redshift range
$0.5\la z\la1.3$. Specific attention is dedicated to the
properties and evolution of the red, quiescent galaxy population
residing along the RS, investigating the impact of different selection
criteria on the robustness of evolutionary trends that are derived
from both the LF and CMR.

The paper is organised as follows. In Section~\ref{data} we give an overview of the data 
and the sample selection used for this work. We address in detail various
incompleteness tests that were considered as well as the derivation of
individual galaxy rest-frame properties and galaxy types. The CMR for VIPERS is
presented in Section~\ref{CMR} and the different selection procedures for
passive galaxies are described in Section~\ref{RS}. The evolution of the 
RS galaxy population is explored in Section~\ref{ev}. The LFs for VIPERS 
galaxies are described in Section~\ref{LF}. Section~\ref{eev} compares
our observational results to predictions of stellar population synthesis 
models. A discussion and the implications of the results for the formation and 
evolution of the galaxy populations is given in  Section~\ref{dis} and 
our main results are summarized in Section~\ref{sum}.

Throughout the paper, we assume a concordance cosmology with 
cosmological parameters of $\Omega_m=0.25$, $\Omega_\Lambda=0.75$, 
and a Hubble constant of $H_0= 100\,h\,{\rm km\,s^{-1}\,Mpc^{-1}}$ with 
$h=0.7$. Unless otherwise stated, the Johnson-Kron-Cousins filter system 
\citep{JM1953} is used. To simplify a comparison with 
previous works, magnitudes and colours are given in the Vega system.


\section{Data}\label{data}

\subsection{Photometric Data}\label{phot}

The optical photometric catalogue is based on $u^{*} g'r'i'z'$ data from the 
T0005 release of the Canada-France-Hawaii Telescope Legacy Survey
(CFHTLS)\footnote{CFHTLS web site: http://www.cfht.hawaii.edu/Science/CFHTLS/}.
The data were collected with the 3.6m CFH optical/infrared telescope on the 
Mauna Kea summit. The four independent contiguous Wide (CFHTLS-W) patches
cover between 25 to 72 deg$^2$ resulting in a total area of $\sim$155 deg$^2$,
of which VIPERS targets two CFHTLS-W fields, W1 and W4.
The final CFHTLS-W photometric catalog reaches in the optical filter bands
80\% completeness limit in AB for point sources of $u^{*}$$=$$25.2$, 
$g'$$=$$25.5$, $r'$$=$$25.0$, $i'$$=$$24.8$, $z'$$=$$23.9$ \citep{T0005,T0006}.
For all bands, total apparent magnitudes were 
measured in \texttt{Sextractor} using \texttt{mag\_auto} in Kron-like
\citep{Kro80} elliptical apertures as measured from the $i'$-band image 
with a minimum Kron radius of 1.2 arcsec (equal in all other bands).
Apparent magnitudes were corrected for Galactic extinction using the 
\textit{COBE} dust maps by \citep{SFD98}, with a median extinction
of $E(B-V)$$\sim$0.025~mag in W1 and $\sim$0.05~mag in W4.
For VIPERS objects with $i'_{AB}$$<$$22.5$ at $0$$<$$z$$<$$1.0$, the CFHTLS T0005 photometric
redshifts have a 1-$\sigma$ uncertainty of $\sigma_z/(1+z)=0.045$, whereas
in the redshift range $1.0$$<$$z$$<$$1.5$ the uncertainty is $\sigma_z/(1+z)\sim0.090$
\citep{CIK09,SGF11}. For galaxies at $0.4$$<$$z$$<$$1.3$, 
typical errors in the observed $u^{*} g'r'i'z'$ CFHTLS photometry that include
all measurement uncertainties (e.g., zero-point and absolute photometric 
calibration, tile-to-tile offset, etc) are of the order 
$\langle\sigma_{u^{*} g'r'i'z'}\rangle=0.038\pm0.029$.
There is a multitude of ancillary data available for the two VIPERS fields which
will be explored in a series of future papers. In the following, we limit the
discussion to the relevant photometric data used for the present analysis. 
Both W1 and W4 are covered at NIR wavelengths by various
photometric surveys. A dedicated WIRcam $K_s$ follow-up survey of the VIPERS
fields (Arnouts et al. 2014, in preparation) was conducted for $\sim$80\%
of the area in W1 and $\sim$96\% in W4 with a 5$\sigma$ completeness level for 
point sources of $K_s$$\sim$$22.0$~mag (AB). 
In W1 and W4 23,759 and  27,371 VIPERS objects have a $K_s$ counterpart,
respectively. Further, 4,591 objects ($\sim$15\%) in W4 that have no
WIRCAM $K_s$-band photometry are covered by UKIDSS-DR9 or, when available,
UKIDSS-UDS-DR8 $YJHK$ data. For 1004 and 1500 spectroscopic sources in
W1 and W4, respectively no $K$-band data is available. However, only 801 out
of these 2504 VIPERS galaxies without $K$-band photometry and 
$2\leq z_{\rm flg}\leq 9.5$ are located at $0.5$$<$$z$$<$$1.2$. 
Based on consistency tests with respect to
uncertainties originating from the photometric calibration (e.g., zero-point
variations or sky-subtraction errors), we conclude that the
$K_{\rm WIRCAM}$ photometry is more reliable than the $K_{\rm UKIDSS}$ data
\citep[e.g.,][hereafter D13]{Dav13}. Thus, when both $K$-band data are
available, we decided to keep the former and neglect the latter.

Through a collaborative effort, we have supplemented our multi-wavelength
photometry with deep GALEX observations in the FUV (1350-1750\AA) and NUV
(1750-2800\AA) (Arnouts et al. 2014, in preparation). A total of 
$T_{\rm exp}>30$~ksec/field yields a 90~\% completeness level of extended 
sources at $24.50$~mag (AB) in both FUV and NUV bands.
Overall, 18,838 sources ($\sim$63\%) in W1 and 3,688 ($\sim$13\%)
objects in W4 have a NUV counterpart. In W1 and W4 there are 4,528 
galaxies ($\sim$15\%) and 1,552 objects ($\sim$5\%) 
with an associated FUV flux, respectively.

\subsection{VIPERS}\label{vip}

The VIMOS Public Extragalactic Redshift Survey 
(VIPERS)\footnote{VIPERS web site: http://www.vipers.inaf.it},
is an ongoing ESO Large Programme designed to measure in detail the spatial 
distribution, clustering, and other statistical properties of the 
large-scale distribution of $\sim$$10^5$ galaxies
across all galaxy types down to $i'_{AB}$$<$$22.5$.
In addition, this survey also measures the physical parameters of 
galaxies, like stellar mass or star formation activity, over
an unprecedented volume of $5 \times 10^{7}$ h$^{-3}$ Mpc$^{3}$ at
$0.5$$<$$z$$<$$1.2$ \citep{GSG13,GuzMsngr13}. The VIPERS sample selection is
based on accurate optical 5-band CFHTLS photometry, combined
with a simple and robust pre-selection in the $(g'-r')$ vs $(r'-i')$ 
colour-colour plane to efficiently remove galaxies with $z$$<$$0.5$ 
\citep{GSG13}. Spectroscopic observations have been collected
with the VIsible MultiObject Spectrograph (VIMOS; 
\citep{LFVIMOS00,LFVIMOS02} at the ESO Very-Large-Telescope (VLT) in multi-object-spectroscopy (MOS)
mode using the low-resolution red (LR-Red) grism 
($\lambda_{\rm blaze}=5810~\AA$) at moderate resolution ($R=210$, 1$''$ slit).
This gives a wavelength coverage of $5500$$<$$\lambda_{\rm eff}$$<$$9500$~\AA\
with internal dispersion of 7.15~$\rm{\AA}$~pix$^{-1}$ \citep{Sco05}.
The combination of our target selection with an efficient,
and aggressive observing strategy using shorter slits \citep{Sco09ESOMsg},
that doubles the multiplexing of VIMOS MOS mode
while at the same time keeping the problematic effects of fringing patterns
($\lambda_{{\rm obs}}$$\ga$$8100$~\AA) under control, allows us to double the galaxy
sampling rate in the redshift range of interest ($\sim$40\%) when compared to 
the sampling of a purely magnitude-limited sample. 

The data set used in the present study is based on the VIPERS 
PDR-1, which was made public in Fall 2013 \citep{PDR1}. 
The survey description, together with its motivation, the sample 
selection and the spectroscopic observations are presented
in \cite{GSG13}. The VIPERS data reduction pipeline 
(Easylife), the survey database system and its tools
are described in \citep{GPS12}. There are several complementary scientific 
investigations using the VIPERS PDR-1 \citep[D13,][]{BMG13,dLTGP13,Mal13,MBB13}.
Previous works include a CFHTLS power spectrum analysis \citep{GGC09}
and a principal component analysis of VIPERS spectra \citep{MGG13}.

The VIPERS PDR-1 contains all spectroscopic observations in the VIPERS
database until the end of the 2011/2012 observing campaign and covers 
an effective area of $\sim$10.32 deg$^2$ which is 
61.3\% of the planned total area ($\sim$24 deg$^{2}$).
For 55,358 objects (93.8\% of total, excluding non-extracted spectra), the
automatic redshift measurement software tool EZ \citep{GFF10} was 
able to derive an estimate of the spectroscopic redshift. All automatic
measurements were validated (or corrected, if necessary) via visual inspection
by two VIPERS Team members independently, who also assigned a quality flag
$z_{\rm flg}$ (i.e., a confidence level) to the final redshift measurement.
In total, 53,608 galaxy redshifts have been measured. 
Finally, a comparison between the spectroscopic and the photometric redshift
measurements was carried out, storing the result in the quality flag decimal
value. The spectroscopic redshift measurement accuracy, reliability and flag 
classification system have been extensively tested and verified through repeated
observations of 1215 galaxies \citep{SGF11,GFF10,GSG13}.

\subsection{Rest-frame Galaxy Properties}\label{rest}

All spectrophotometric rest-frame properties of the VIPERS galaxies were
derived using the SED fitting
program Hyperzmass, an updated version of the photometric redshift
code Hyperz \citep{BMP00,BKP10}.
For the spectroscopic PDR-1 catalogue, the dust content of the galaxies was
modelled with both the Prevot-Bouchet extinction law, that is based on the 
Small Magellanic Cloud dust properties \citep{Prev84,Bou85}, and an 
extinction relation calibrated using starburst galaxies \citep{Cal00}. The 
choice between the two extinction relations was performed on the basis of the
smallest derived $\chi^2$ value. The final extinction magnitudes range 
between $A_V=0$ (no dust attenuation) to $A_V=3$ (strong dust attenuation). 

Absolute luminosities were derived using the apparent magnitude that most
closely resembles the observed photometric passband, shifted to the
redshift of the galaxy under consideration. Thanks to our extensive 
multi-wavelength photometry (from $FUV$ to MIR wavelengths), 
the $k$-correction factor was as small as possible and much less sensitive to
the adopted SED template type than using a global filter transformation that is
confined to a single specific filter passband \citep[see appendix~A of][]{ITZ05}.
The following analysis is restricted to rest-frame Johnson $U$, $B$, and
$V$ bands as they are covered across the whole optical to NIR observed filters
over the redshift range of our main interest.
Typical uncertainties in the SED fitting process that include 
measurement uncertainties in the zero-point, SED extrapolation,  
and adopted template libraries, range, for example, in the $B$-band from 
$\sigma_B=0.04$ ($0.4$$<$$z$$<$$0.7$) to $\sigma_B=0.05$ ($0.7$$<$$z$$<$$0.9$) and 
$\sigma_B=0.07$ ($0.9$$<$$z$$<$$1.1$) to $\sigma_B=0.14$ ($1.1$$<$$z$$<$$1.3$).
These uncertainties are included in the final error budget and are used 
in the derivation of the intrinsic scatter of the CMR (see Section~\ref{sca}).
In this work, we focus on exploring the differences between the two main 
populations of passive and star forming galaxies using the rest-frame
$(U-V)$ colours. For the $U$-band we adopt the original $U_J$-Johnson
filter \citep{JM1953}; hereafter $U$-band), which is particularly 
sensitive at blue wavelengths and allows a precise distinction between red 
and blue galaxies. One advantage is that the $U$-band filter response traces
the 4000\,\AA\ break and therefore the $U$-band is a good proxy for the 
stellar population age of galaxies \citep{Bru83,Kau03}.
In Appendix~\ref{uband} we discuss our filter choice and provide a comparison
to other $U$-band filters used in the literature.

%
%
\begin{figure}
 \centering
 \includegraphics[width=0.5\textwidth]{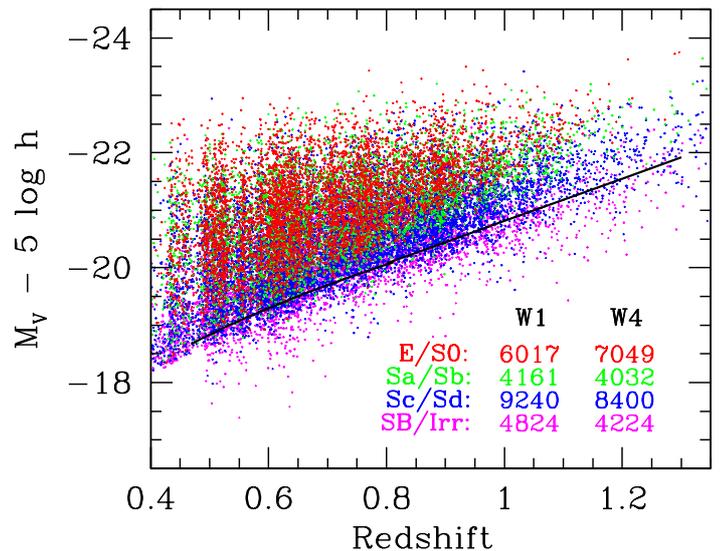}
 \caption{Luminosity-redshift relation for galaxies in the VIPERS PDR-1.
  Galaxies are colour-coded with respect to their SED type classification. 
  The solid line indicates the selection boundaries for 90\% completeness.}
 \label{Mvz}
\end{figure}

\subsection{Spectrophotometric Type Classification}\label{type}

Spectrophotometric galaxy types were derived by fitting the magnitudes
with a small set of spectral templates as described in \citep{IAMC06}.
In particular, the six reference templates consist of four locally
observed spectral-energy-distributions (hereafter SEDs) 
\citep{Col1980} and two starburst SEDs from \cite{Kinney1996}. All 
templates were first individually extrapolated to UV and MIR wavelengths, then 
pairs of templates were interpolated to create a final set of 62 synthetic SEDs,
and finally those were optimized with {\em Le Phare} \citep{SE1999MN,IAMC06}. 
As our templates are constructed from real observed 
galaxies rather than synthetically generated templates, they already contain 
the typical amount of dust present for that given spectral galaxy class. The 
whole spectroscopic sample is then classified into four galaxy types,
corresponding to early-type spheroids and spiral bulge-dominated galaxies 
(elliptical, lenticular and S0/Sa, hereafter E/S0), early-type spiral galaxies 
(hereafter Sa/Sb), late-type spiral galaxies (hereafter Sc/Sd), and starburst
and irregular galaxies (hereafter SB/Irr).

Figure~\ref{Mvz} shows the luminosity-redshift relation of VIPERS.
Galaxies are divided into their respective SED type classes.
Because the flux intensity decreases with the square of the 
luminosity distance ($S=L/4\pi\,D_{\rm L}^2$), with increasing
redshift any galaxy sample will be incomplete at faint luminosities. The solid
line denotes the selection boundaries for a sample completeness of 90\%, 
which was derived as in \cite{MBB13}. 

In our subsequent analysis, the SED type classification serves as primary
reference against which the completeness and contamination levels of each
independent selection criteria of early-type galaxies are compared to. 
The pros are that for the SED type assessment
the complete optical to NIR multi-wavelength information can be applied.
The  broader wavelength coverage reduces the number of possible 
contaminations from special objects for any given galaxy type. Our concept
of using the SED type classification as a reference for selecting
early-type galaxies gets independent support from a recent zCOSMOS
study \citep{MPC13}.

%
%
\begin{figure}
 \centering
 \includegraphics[width=0.5\textwidth]{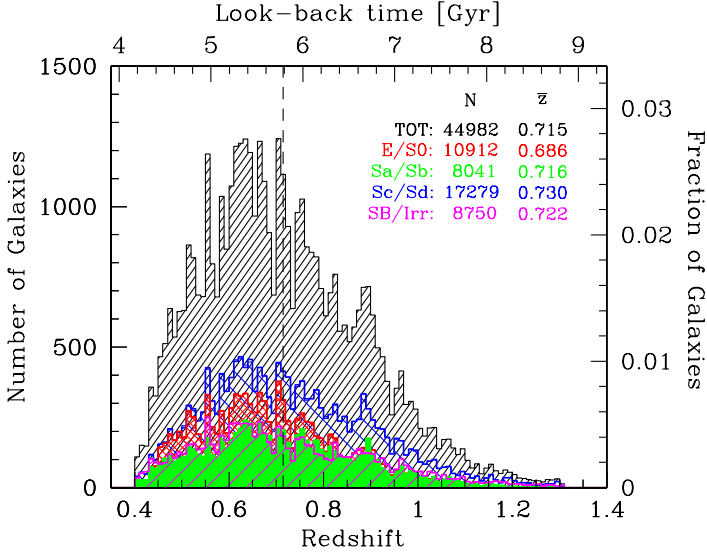}
 \caption{Redshift distribution for the VIPERS PDR-1 sample used in the
  present study. Only galaxies with $2\leq z_{\rm flg}\leq 9.5$ are considered.
  The mean of the total distribution is indicated with a dashed line. The total
  number of galaxies and mean redshift are given for each galaxy type.}
 \label{zdis}
\end{figure}

\subsection{Sample Selection and Redshift Distribution}\label{sel}

The sample used for the present work is composed of 44,982 galaxies 
($N_{\rm W1}=23,210$ and $N_{\rm W4}=21,772$) with $2\leq z_{\rm flg}\leq 9.5$
(corresponding to a confidence level $>$95\% in the redshift measurement)
covering the redshift range between $0.40<z<1.3$ and look-back times of
$\sim$4.2--8.7~Gyr. An additional 763 objects with $z<0.4$ and 244 galaxies
with $z>1.3$ exist with the same quality flags, but have been excluded due to
sample incompleteness. The redshift distribution of these data is presented in 
Figure~\ref{zdis}. The mean redshift is $\overline{z}=0.715$ 
(indicated with a dashed line), and the median is 
$\langle z\rangle=0.694$. Figure~\ref{zdis} also shows the redshift 
distribution of the different galaxy SED types. Using repeated observations of
1215 galaxies with $z_{\rm flg}\in{2,3,4}$ (excluding stars and AGNs), the 
typical (root-mean-square) uncertainty in the spectroscopic redshift 
measurements is $\delta_z/(1+z)=0.00047$, which corresponds to a radial 
velocity accuracy (1-$\sigma$ scatter) for individual VIPERS measurements of 
$\delta_v=141(1+z)$ km sec$^{-1}$.

\begin{table}[t]
\caption{Percentage of early- and late-type galaxies in the 
parent and spectroscopic VIPERS sample.}
\label{zmc}
\centering                         
\begin{tabular}{c c c c c c c}   
\hline\hline
Type & \multicolumn{2}{c}{$0.4<z<0.5$} & \multicolumn{2}{c}{$0.5<z<0.6$} & \multicolumn{2}{c}{$0.6<z<1.0$} \\
     & Parent & Spec & Parent & Spec & Parent & Spec\\
\hline
E/S0   & 22.0 & 27.1 & 24.9 & 27.4 & 26.7 & 24.3 \\  
Sa/Sb  & 20.3 & 18.5 & 19.1 & 16.3 & 20.6 & 18.3 \\  
Sc/Sd  & 42.1 & 32.6 & 36.0 & 36.9 & 35.4 & 38.9 \\  
SB/Irr & 15.7 & 21.8 & 20.1 & 19.5 & 17.3 & 18.5 \\  
\hline                        
\end{tabular}
\end{table}

\subsection{Selection Function}\label{biases}

In order to draw robust conclusions from the observed distribution of physical
parameters in the VIPERS galaxy sample, we have to consider the
influence of possible systematic effects which might enter our
spectroscopic sample. In particular, a potential bias may arise if galaxies
with extreme (e.g., colour) properties are under- or over-represented in the 
final spectroscopic sample. Such systematics could
be either due to incompleteness in colours of the parent photometric sample,
or a variable redshift measurement success rate.

\subsubsection{Spectral Type and Colour Completeness}

One potential source of bias in our spectroscopic sample might be the
systematic deficit of a specific class of galaxies, in particular early-type
galaxies. The origin of such a selection effect might be twofold. Brighter
early-type galaxies with large sizes are preferably located in dense
environments and display a stronger clustering compared to their late-type
counterparts. Furthermore, the number density of fainter and smaller late-type
galaxies increases with increasing redshift. Because of MOS mask design 
restrictions, spectroscopic surveys will target only a fraction of the
total galaxy population and usually select against early-type galaxies due to
their increasing relative physical underabundance with redshift and their 
stronger clustering properties. 
However, the parent target sample is sufficiently sparse that the physical
effect of clustering is small and can be largely accounted for during the 
automatic slit assignment procedure.
In quantitative studies of clustering this effect can be corrected to within a
few percent through a proper weighting scheme based on the angular correlation
functions of the observed and full target samples.

Another potential reason for a deficit of early-type galaxies might be a
lower efficiency in the redshift measurement process due to the lack of
prominent emission lines in the spectra.  
To assess this possible systematic uncertainty, we compare our spectroscopic
sample to the total $u^{*}g'r'i'z'$ photometric magnitude-limited CFHTLS catalog
in the VIPERS fields \citep{CIK09} for which photometric redshifts could
be obtained using {\em Le Phare} \citep{SE1999MN,IAMC06}. For
the parent sample, the same SED modelling process and photometric type
classification has been conducted as described for the spectroscopic sample 
in Section~\ref{type}. The parent sample consists of $N$$=$$345,605$ galaxies
with $0.4$$<$$z_{\rm phot}$$<$$1.3$. We note that possible
uncertainties due to photometric redshift errors are small
($\sigma_{z_{\rm phot}}=0.0387(1+z)$, catastrophic outliers $\la$3-4\%
for $z_{\rm flg}\in{2,3,4}$; \citep{CIK09}. In
Table~\ref{zmc} we compare the percentage of photometrically classified
early-type (E/S0) and late-type (Sa to Irr) galaxies in the parent and in the
spectroscopic sample. The fractions of each galaxy type are given for 
different redshift intervals probed by VIPERS.
Comparing the fraction of early-type galaxies in the parent and spectroscopic
datasets we deduce that the incompleteness of the early-type population
in the spectroscopic sample is negligible. On average, early-type galaxies
are not deficient in the total spectroscopic sample. Only for the
highest-redshift part ($1.0<z<1.3$) E/S0s appear to be slightly deficient
by $\sim$4.3\%. Overall, we find no significant dependence of the redshift
completeness on any particular galaxy type.

%
%
\begin{figure}
 \centering
 \includegraphics[width=0.5\textwidth]{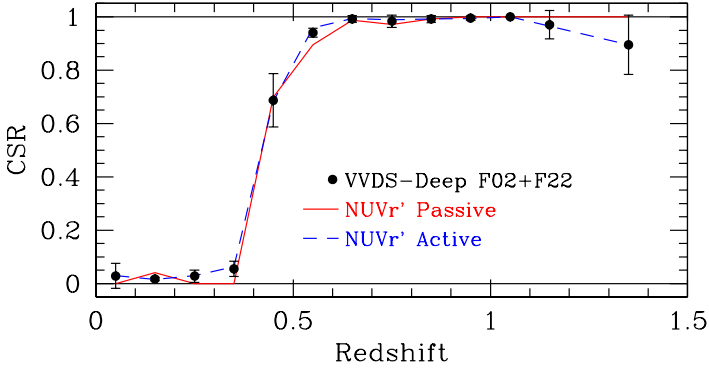}
 \caption{Colour completeness for VIPERS. The sample is complete in colours
  between $0.5<z<1.3$, in the sense of including a proportionally
  correct fraction of galaxies with different colours.
  A $NUVr'$ selection to separate (red) passive (solid line) from 
  (blue) active, star-forming (dashed line) galaxies shows that the CSR 
  is independent of galaxy type.}
 \label{CRS}
\end{figure}

To establish the completeness in galaxy colours for our sample, we have
estimated the effect of the varying sampling rate using spectroscopic samples
from the purely magnitude-limited  VIMOS-VLT Deep Survey (VVDS,
\citep{VVDS05,VVDS13}. We selected galaxy samples from the VVDS F02
and F22 fields applying an identical colour selection criterion as the one used
in VIPERS, and then derived the colour completeness (see Figure~\ref{CRS}). We
define the Colour Sampling Rate (CSR) as ${\rm CSR}(z)=N_{\rm VL}/N_{\rm TOT},$
which is the ratio of the number of galaxies in a VIPERS-like selected sample
$N_{\rm VL}$ to the total number of galaxies $N_{\rm TOT}$ in the same
redshift bin. Above $z=0.6$, the CSR is independent of the redshift with 
CSR$\sim$1 up to $z\sim1.2$ and the VIPERS sample is equivalent to a purely
magnitude-limited sample with $i'_{AB}$$<$$22.5$ \citep{GSG13}.
If the VVDS sample is split into passive (red) and active, star forming
(blue) galaxies using a $(NUV-r')-(r'-K)$ selection (see Section~\ref{nuv}),
the completeness level remains at a similar level, as demonstrated in
Figure~\ref{CRS}. Below $z=0.6$, a lower completeness for passive galaxies 
is somewhat expected due to the sharp cut in the colour selection.
However, we find only a minor decrease in the CSR of passive 
galaxies. At $z=0.6$, passive galaxies have a completeness of $\sim$94\%,
whereas blue galaxies are complete at $\sim$97\%. At $z$$\sim$$0.5$, the CSR is 
$\sim$82\%, whereas at $z$$\sim$$0.45$ it is $\sim$69\%, independent of 
the galaxy type. 
Significant biases arise when the colour completeness is $\lesssim$50\%
\citep{Fra07}. As for $z$$<$$0.45$ the CSR is still $\sim$69\%,
our  sample contains a large fraction of the galaxy population between
$0.4$$<$$z$$<$$0.5$ which should not be significantly affected by selection
effects. In any case, in the subsequent analysis this redshift range is treated
separately from the rest of the sample. Overall, we conclude that across the redshift
interval probed in this work we detect no significant dependence of the CSR
on galaxy type.

\subsubsection{Statistical Weights}\label{swei}

VIPERS is based on a robust colour selection criterion in the 
$(r'-i')-(u'-g')$ colour-colour plane to effectively target galaxies at
$z$$>$$0.5$ \citep{GSG13}. The completeness of the survey can vary
as a function of different observed quantities, like galaxy redshift, 
magnitude, and colour, and it also varies for each targeted VIMOS quadrant.
Here we describe the tools that allow us to evidence and correct
for incompleteness effects in our analysis of the colour magnitude relation
(Section~\ref{CMR}), the luminosity function (Section~\ref{LF}) and the GSMF 
(D13). A different application of the statistical weights
with respect to galaxy clustering is given in \cite{dLTGP13}.

We define the Target Sampling Rate (TSR) as the fraction of all observed 
spectroscopic sources with respect to the total photometric sample
${\rm TSR}(i'_{AB})=N_{\rm spec}/N_{\rm phot},$
where $N_{\rm phot}$ and $N_{\rm spec}$ are the number of potential targets
present in the parent photometric catalogue and of observed
spectroscopic sources, respectively. The TSR varies only 
as a function of apparent magnitude.

The Spectroscopic Success Rate (SSR) can be computed as 
${\rm SSR}(i'_{AB},z)=(N_{\rm spec}^{\rm gal}-N_{\rm spec}^{\rm fail})/
N_{\rm spec}^{\rm gal},$
where $N_{\rm spec}^{\rm gal}$ is the number of spectroscopically observed
galaxies (excluding broad line AGNs, and spectroscopically identified stars).
$N_{\rm spec}^{\rm fail}$ is the number of sources
without a reliable redshift confirmation  (i.e., ``failures'' with 
$z_{\rm flg}=0$, or low confidence with $z_{\rm flg}=1$). The
SSR depends not only on the observation details (i.e. the VIMOS quadrant)
and on the redshift of the objects, but also on their apparent magnitude,
as the spectroscopic signal-to-noise ratio ($S/N$) decreases with 
increasing magnitude.

Finally, unobserved sources in the survey are corrected by using a
statistical weighting scheme associated to each galaxy $g$ with a secure
redshift measurement \citep{ITZ05} , where $w_g^{\rm TSR}$, 
$w_g^{\rm SSR}$ and $w_g^{\rm CSR}$ are the inverse of the TSR, SSR and 
CSR, respectively. The total survey completeness weight $w^{\rm TOT}$ is given 
as $w_g^{\rm TOT}=w_g^{\rm TSR}\times\ w_g^{\rm SSR}\times w_g^{\rm CSR}$.
This weighting scheme corrects for both sources that were not observed or
unidentified (i.e., objects with unknown redshifts because of poor spectral
quality). Since the VIPERS spectroscopic targets were randomly selected from
the parent sample, the TSR is independent of the apparent magnitude, being
stable at $\langle$TSR$\rangle \gtrsim$$40\%$. In contrast, the SSR is a
function of the redshift and the $i$-band selection magnitude, which is
connected to the $S/N$ of the spectrum. Hence, the SSR ranges from
$\sim$95\% at $i'_{AB}\sim20.2$ to $\sim$75\% at 
$i'_{AB}\sim22.2$ \cite[see also][]{dLTGP13,GSG13}.

The weighting scheme was applied to all observed galaxies apart from two
special object classes. First, as the spectroscopic catalogue contains a small
fraction of $X$-ray detected sources (``compulsory'' targets), the TSR for 
these objects is higher than the global one and therefore was computed
separately. Secondly, the contribution of these ``compulsory'' targets to the
total sample is only $\sim$2\%. In some cases, a secondary spectrum 
was detected within a slit assigned to a target. For these serendipitous 
sources the SSR would be much lower than the overall SSR as the majority of
these objects are either faint or not well centred within the slit,
resulting in low $S/N$ spectra. These secondary objects have not been included
in the PDR-1 and will be analysed in the future.


%
%
\begin{figure*}
 \centering
 \includegraphics[width=1.0\textwidth]{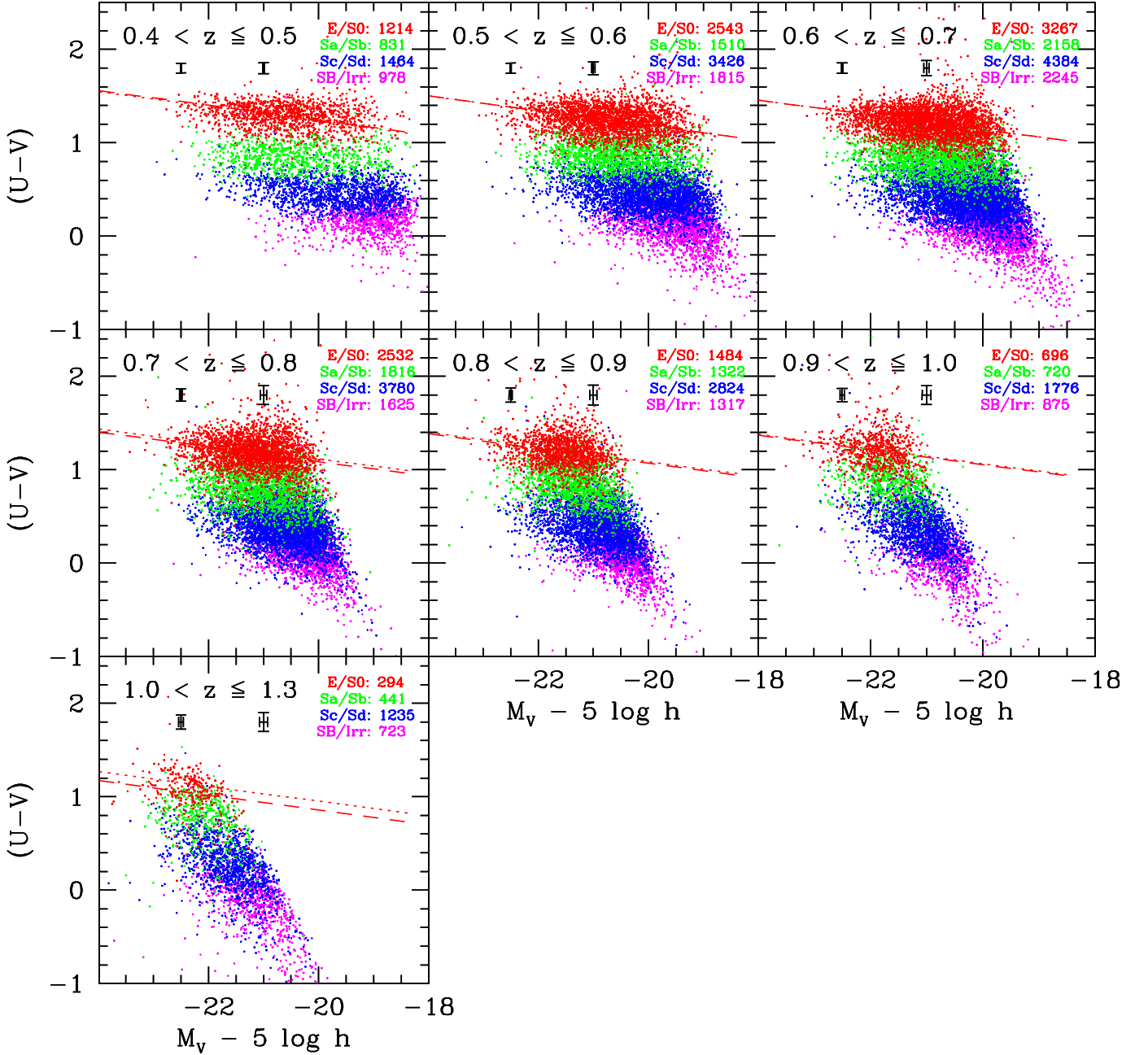}
\caption{Evolution of the Colour-Magnitude Relation for the VIPERS PDR-1 
 sample from $z=0.4$ to $z=1.3$. Galaxies are colour-coded with respect 
 to their derived SED type. The number of galaxies for a specific SED type 
 is shown in each slice. The lines in each panel indicate the best-fitting CMR
 for galaxies on the RS within the given redshift interval: 
 classical approach (dotted), SED galaxy types (dashed). The fit for
 the colour-bimodality method is not shown as it coincides with the dashed line.
 Typical error bars (black) are shown in each redshift slice.}
 \label{cmra}
\end{figure*}

\section{The Colour-Magnitude and Colour-Stellar Mass Relations in VIPERS}\label{CMR}

%
%
\begin{figure*}
 \centering
 \includegraphics[width=1.0\textwidth]{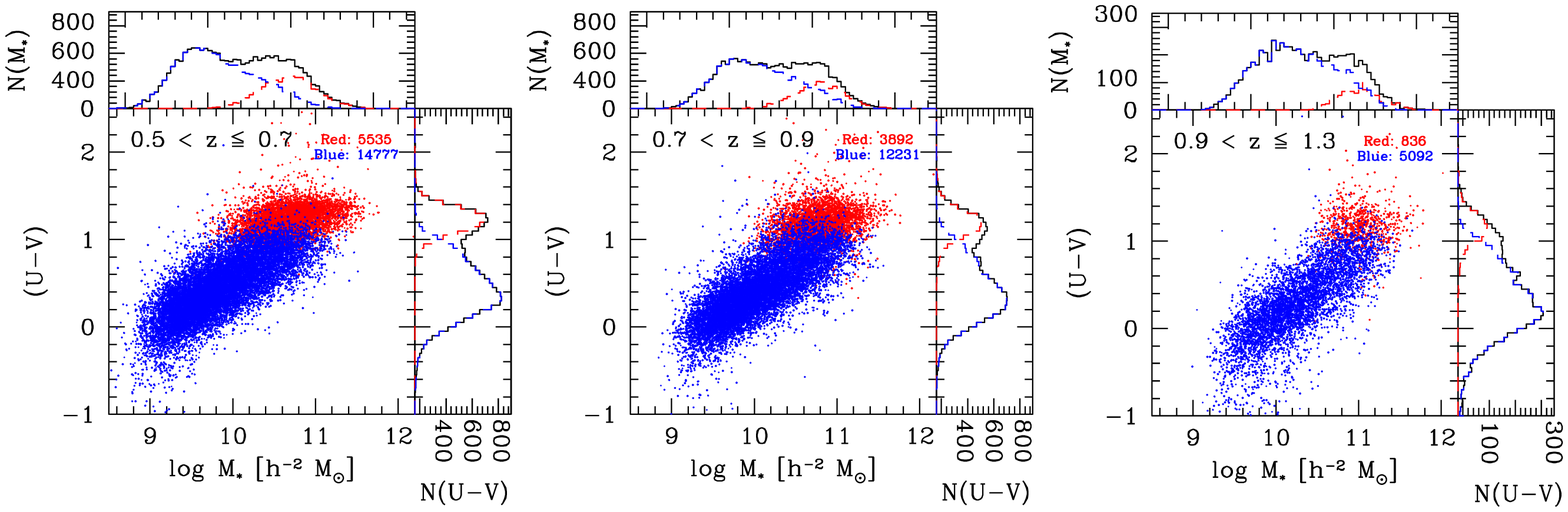}
\caption{The relation between galaxy colour and stellar mass for the
  VIPERS PDR-1 sample, within three broader redshift bins. This plot
  confirms that the most luminous blue galaxies observed in the CMR
  are in fact very massive objects. In particular, we note how (a) the
  highest-mass edge of the blue cloud evolves with cosmic time, with a few
  objects as massive as $\log(M_{\star}/M_{\odot})\sim 11.5$
  displaying colours typical of star-forming galaxies;
  (b) conversely, the corresponding highest-mass edge of the RS is
  substantially frozen over the explored redshift range. }
 \label{csm}
\end{figure*}

Figure~\ref{cmra} shows the evolution of the rest-frame $(U-V)$ vs. $M_V$
colour-magnitude relation (CMR) for the VIPERS PDR-1 sample, divided into 
seven redshift slices between $z=0.4$ to $z=1.3$. Galaxies are colour-coded 
according to their derived SED type classification as described in 
Section~\ref{type}. Thanks to the large size of the survey, all redshift bins 
are well populated by a significant number of galaxies for each individual SED
type, allowing a careful description of their relative evolutionary 
patterns.  This is possible because of the precision of the CFHTLS 
photometry, as shown by the plotted colour error bars for two reference 
luminosities, $M_V-5\ {\rm log}\ h=-23$  and $M_V-5\ {\rm log}\ h=-21$.
The CMR covers a range of about four magnitudes in the first redshift bin,
which at $z=1$ is reduced to 2.5 due to the apparent magnitude cut of the sample.
The ``blue cloud'' is the part of the diagram that is less affected by
such incompleteness at the faint limit (as shown in Figure~\ref{Mvz}).  
Keeping these limitations in mind, a number of important trends emerge 
from the CMR plot. First, we note that the rest-frame colours of all
galaxies become bluer with increasing redshift; this happens for all
SED types.  This trend cannot be the effect of a bias induced by the 
survey apparent magnitude cut. As demonstrated in Section~\ref{swei}, 
our completeness in colour remains at a very high level ($>$90\%) across the
whole redshift range, with variations from one bin to another of only 
$\sim$5\%. Therefore, for a given redshift slice, our sample is expected to be 
representative of the correct galaxy population mix present at that
look-back time among relatively luminous galaxies. A second relevant 
feature of the diagrams is that at fixed colour, the upper luminosity limit
of the distribution is higher at higher redshifts.

We can also precisely identify and measure the position of the 
RS and trace its evolution up to redshift $z\simeq1$. In particular, the bulk 
of the red luminous population (hence the
most massive objects) is already in place at $z\sim1$.  
This observational evidence agrees with findings from deeper, smaller-volume 
surveys \citep[e.g.,][]{Bel04b,Cim2006,Bun07,Fab07,ISLF10}.
However, for $z>0.7$ we note the presence of a number of
super-luminous red galaxies, with $M_V-5\ {\rm log}\ h<-23$.  It is
clear that these are very rare objects, which are detected in VIPERS
thanks to its unprecedented volume at these redshifts. The presence of 
these objects on the RS is a clear indication for both a high-redshift
formation  epoch for the bulk of the stellar populations and an early 
stellar mass assembly age for these systems, contrary to speculations that 
the two ages could be decoupled and therefore end up being rather different
from one another \citep{BCF96,DF01}.

Furthermore, the ability of VIPERS to push deep into the bright end
of the luminosity function also has a beneficial effect on the blue
population. As we see from the same figure, a large number of green or blue
very luminous galaxies is detected at all redshifts sampled by VIPERS, 
in particular in the highest redshift bins, where we detect galaxies 
with $M_V-5\ {\rm log}\ h<-23$ and very blue colours. 

Clearly, the crucial question in terms of evolutionary patterns is
whether these are objects of normal mass, experiencing an extremely
active star-burst phase, or are rather massive objects going through a
normal star formation phase without a strong starburst. Our results on the 
stellar mass function of red and blue galaxies in VIPERS (D13),
show that at $z\sim1$ one finds comparable abundances of blue and red galaxies 
with masses as large as $\log(M_{\star}/M_{\odot})\sim 11$. 
The majority of these luminous blue galaxies present undisturbed 
morphologies without signs of interactions, tidal features, or close 
companions and have no strong $X$-ray emission. This suggests that a 
large fraction of the very luminous blue objects are in fact also very 
massive systems. To verify this more explicitly, we make use of the
stellar masses computed as described in D13, to construct 
the corresponding $(U-V)$ colour-mass diagram of our sample.  This is shown in 
Figure~\ref{csm}, split over three broad redshift bins.  
The figure confirms that the most luminous blue galaxies correspond 
in fact to truly massive objects that are still forming stars.  
Specifically, at higher redshifts ($z>0.7$) we detect a higher fraction of  
massive galaxies with fairly blue colours (as massive as
$\log(M_{\star}/M_{\odot})\sim 11.5$) compared to the lowest 
redshift interval at $0.5<z\leq0.7$.  It is quite natural
to conclude that these objects are today located on the red sequence:
this ``migration'' is in fact apparent when comparing the three
diagrams, together with the overall evolution in the number density of
blue and red galaxies in different mass bins (D13,
Fig. 14). Figure~\ref{csm} demonstrates that the build-up of the red envelope 
of the RS (the highest-mass edge) has been completed at $z\sim1.1$ 
and remains constant over the whole explored redshift range
($\log(M_{\star}/M_{\odot})\sim 11.8$), but the 
overall population of the RS for masses 
$\log(M_{\star}/M_{\odot})> 10.5$ 
increases by a factor of $\sim$2.4.  We argue this
represents substantial evidence for a quiet build-up of the main
body of the RS, without the need of major dry merger events.  
\cite{IMCLF13} showed that the overall measured star formation at $z<1$
is not efficient enough to produce a population of massive star-forming galaxies. 
The significant population of massive blue galaxies that we have
found to be already in place at $z\sim1$, makes the formation of new 
massive blue galaxies at lower redshifts not necessary. A more
detailed study of the observed evolution of the bright/massive edge
of the ``blue cloud'' will be the subject of a future paper
(Scodeggio et al., in preparation).

\section{The Red-Sequence}\label{RS}

An important point in all discussions about the evolution of early-type
galaxies is how to select a sample of these galaxies that is sufficiently
pure and complete at the same time. In this work we compare different selection 
procedures to construct a sample of red, old and passive galaxies of high 
purity, using several independent criteria to separate these galaxies
from blue, star-forming ones, and we also explore the possible effects of dust 
obscuration in high-redshift, dusty red galaxies which are located around 
the RS and within the ``green valley''.

\subsection{The Classical Approach}\label{class}

One popular approach to distinguish the red and blue galaxy population
is to use a {\em fixed cut in galaxy colours, not-evolving with redshift}
\citep[e.g.,][]{Bel04b,Fra07}. When using this method, 
many works adopted the slope in the CMR as defined by elliptical galaxies
in the Coma cluster, which is well-known and where selection effects are
under control. There are two main motivations for using this criterion:
(i) straightforward comparison to previous works in the literature, 
(ii) limiting possible uncertainties due to zero-point variations in the 
photometry that arise from the transformation of observed magnitudes
to the rest-frame at different redshifts.

Here we follow the prescription outlined by \cite{Bel04b}, 
which we refer to as the ``classical approach''.
Galaxies that have rest-frame colours $(U-V)>1$, regardless of
their morphological or SED type, are considered as part of the RS.
The evolution of the RS is derived 
by fitting a linear relation between colour and absolute magnitude, 
keeping the slope of the relation fixed to the Coma cluster value 
at all redshifts. As redshift progresses the assumption of a non-evolving
selection cut with redshift becomes problematic as the number density
of early-type galaxies with intermediate mass
($10.8<\log(M_{\star}/M_{\odot})< 11.1$) decreases by a factor of $\sim2.5$
and that of high mass galaxies ($\log(M_{\star}/M_{\odot})\ge 11.4$) by 45\%
from $z=0.6$ to $z=1.2$ (D13). Therefore, the classical
approach will select only the reddest and most luminous (hence also most 
massive) early-type galaxies, which are not representative of the whole 
early-type galaxy population. These massive early-type galaxies have formed
the bulk of their stars at high redshift ($z>2$) and will therefore comprise
quiescent evolved  stellar populations. 
However, any selection method using optical colours only, will contain
a non-negligible contribution of dusty red galaxies 
\citep{Stra01,CDM02,GBD03,Bel04a,Wei05}. To disentangle the contamination of
the early-type galaxy samples due to such effects, we will later on use a 
combination of GALEX $NUV$ and NIR colours (Section~\ref{nuv}).

\subsection{Colour Bimodality}

The most evident aspect of the distribution of galaxy colours is of course
the colour bimodality. In the VIPERS data, as in local surveys, this is a very 
general feature that is evidenced when using different rest-frame $UBVRI$ 
colour combinations, spectral diagnostics such as the 4000\,\AA\ break 
\citep{PDR1} or morphological quantities 
(Krywult et al. 2014, in preparation). In this work we use the $(U-V)$ colour,
as it is highly sensitive to the slope of the blue/ultraviolet continuum,
representing a natural tracer for SF galaxies (see Appendix~\ref{uband}).
In particular, our $U$-band filter represents a good measure of the
overall SF activity in our galaxies. However, similar to the classical
method the colour bimodality does not take into account the contamination 
by AGNs or dust-obscured red galaxies (see Section~\ref{nuv}).

%
%
\begin{figure}[t]
 \centering
\includegraphics[width=0.5\textwidth]{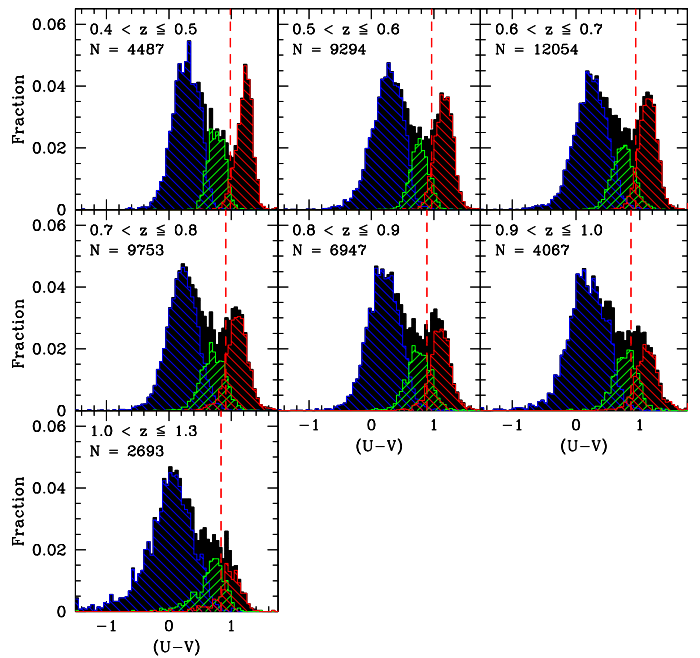}
 \caption{The rest-frame $U-V$ colour distribution in the different
   redshift bins, split into three broad SED galaxy types 
   of blue galaxies (Sc/Sd and SB/Irr, blue histograms),
   green galaxies (Sa/Sb, green), and red (E/S0, red).
   The dashed line in each panel indicates the adopted 
   separation for red and blue galaxies, defined by a fit
   to the local minima of the colour distributions. The colour-bimodality 
   separation is also a good representation of the different SED types of
   early-type and late-type galaxies.}
 \label{bimoSED}
\end{figure}

\subsubsection{Classification and Evolution Since $z\sim1.3$}\label{green}

Another approach to segregate red and blue galaxies is to adopt a 
{\em variable cut in galaxy colours that evolves with redshift}  
\citep[e.g.,][]{WAB09,Pen10}. This appears to be a more physical method to
separate red from blue galaxies than assuming a fixed cut a priori without
accounting for the redshift evolution.

The procedure we adopt is the following: first, we project the colour-magnitude 
relation onto the $(U-V)$ colour axis, 
after subtracting out the slope defined by elliptical 
galaxies in the Coma cluster (see Section~\ref{class} and \ref{fits}).
Figure~\ref{bimoSED} illustrates the rest-frame $(U-V)$ colour
distributions, after subtracting out the Coma slope. A bimodal $(U-V)$ colour
distribution is evident across the whole redshift interval, characterized by
two peaks and a well-defined minimum. The location of these three extrema 
depends weakly on the adopted projection method (i.e., if 
a reference RS slope is subtracted out or not, see Section~\ref{ev}). 
Next, early-type (red) galaxies are separated from late-type (blue) galaxies by 
measuring the local minimum in the colour distribution within each redshift bin. 
For the colour separation value, a simple linear evolution with redshift is 
assumed and by fitting the observed local minima and redshift value pairs,
we derive a separation in the rest-frame $(U-V)$ colour distribution which
evolves as $(U-V)=1.1-0.25\times\,z$,  
denoted with the red dashed line in Figure~\ref{bimoSED}.
This partition separates well the two main populations of red and green-plus-blue 
galaxies. We emphasize that to measure the decrement of passive 
galaxies in the colour projection, the evolution of the green galaxy population 
needs to be considered. For this reason, we divide the rest-frame $(U-V)$
colour distributions in Figure~\ref{bimoSED} into the three different global 
SED type classes, representing the red, green and blue galaxy population. Note
that for this comparison the SED types of Sc/Sd and SB/Irr were combined 
in order to emphasize the broader transition between the galaxy populations
of the green valley and the blue cloud.
Thanks to the very good statistics  of VIPERS, it is possible for the first 
time at these redshifts to separate galaxies in the transition zone from the 
blue galaxy population, although this method is not corrected for possible 
contamination effects by dusty red galaxies.

%
%
\begin{figure}
 \centering
 \includegraphics[width=0.5\textwidth]{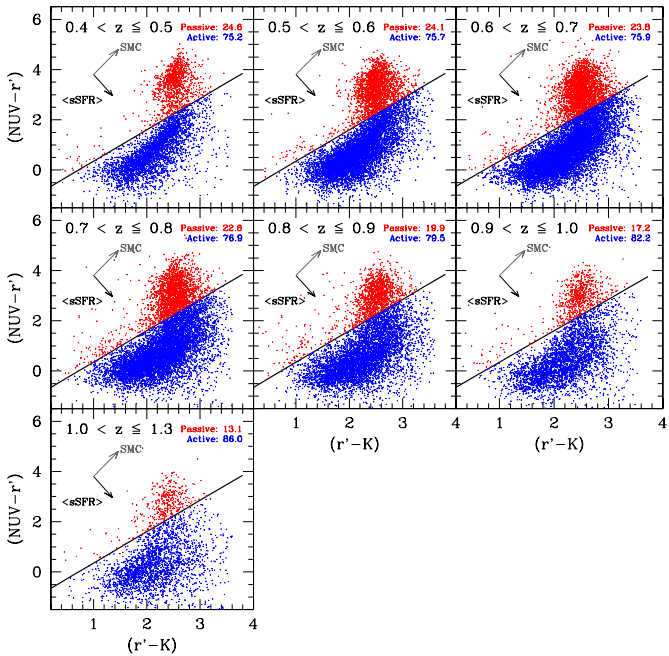}
\caption{Evolution of the rest-frame $(NUV-r')$ vs $(r'-K)$ plane for 
VIPERS. A selection along the lines of of constant sSFRs 
(black $\langle \mathrm{sSFR}\rangle$ arrow) and dust-extinction (grey SMC arrow) gives
a clean separation for the majority of passive (red) and star-forming active
(dusty red and blue) galaxies. The fraction of galaxies (in percent) 
classified as passive and active is given in each panel.}
 \label{nuvr}
\end{figure}

\subsection{NUVr' Classification}\label{nuv}

Red galaxy samples selected on the basis of optical colours contain a 
substantial contamination of $\sim$30-40\% by Seyfert 2 AGNs and/or dust
obscured SF galaxies 
\citep{Stra01,CDM02,GBD03,Bel04a,Gia05,Wei05,Fra07,HGM08,GBW09}. Dust obscuration 
is less significant at low redshift, where $\sim$75\% of red galaxies contain
little amount of dust \citep{GBD03,Bel04a,WGM05} and for
the remaining objects current space-based facilities permit a visual detection
of the dust features up to $z$$\sim$$1$ \citep[e.g.,][]{McI05b,FBZ09,FJSC09,Tas09}.  

However, at high redshift ($z>1$) the presence of dust becomes a major 
complication, with red galaxy samples getting significantly contaminated by
dusty SF galaxies. This affects both the colour measurements and the 
assessment of intrinsic colour variations in the galaxies. Several alternative 
methods have been proposed to identify systems with colours dominated
by internal dust absorption, such as mid-IR photometry \citep{Pap05}, 
an optical-NIR rest-frame colour combination, e.g., $UVJ$ selection 
\citep{W09UVJ}, or visual dust extinction constraints from 
SED modelling \citep{Bra09,Bra11} or a using a $UV$-optical-NIR colour
combination \citep[e.g.,][]{Sal05,AWLF07}.

To explore the effects of possible dust obscuration within our early-type
sample, we introduce a combination of GALEX $UV$, optical and $NIR$ fluxes.
In Figure~\ref{nuvr} we illustrate our selection of passive red galaxies 
in the rest-frame $(NUV-r')\,vs\,(r'-K)$ plane \citep{ALFC13}.
A similar classification into quiescent and SF galaxies
with different star formation activity level can be obtained using 
$(NUV-r')\,vs\,(r'-J)$ colours \citep{ISLF10}, although our addition of the
$K$-band permits a much sharper separation between quiescent and galaxies 
with low SF activity, particularly for dust-obscured red objects at $z>0.8$
with $(r'-J)>0.8$ (AB). Further, $NUV-r'$ colours represent an excellent
indicator for the current-to-past star formation ratio. This ratio decreases
with mass for star forming galaxies \citep{Sal05,AWLF07}.
While the $NUV$ passband traces stellar populations with a mean light-weighted
age $\langle t \rangle\sim 10^8$ yr, the $r'$ band is sensitive to
$\langle t \rangle\ge 10^8$ yr \citep{MFS05} and 
the current to the past averaged SF rates correlates with the birth
parameter $b$ as $b=\mathrm{SFR}(t<10^8)/\langle \mathrm{SFR}\rangle$.
Passive galaxies with de Vaucouleurs light profiles can be associated 
with $NUV-r'\ge1.70$ (Vega) and $b\leq0.1$ \citep{Sal05}.
We then define early-type galaxies (hereafter $NUVr'$ selection)
as all the galaxies which are redder than 
\begin{equation}
(NUV-r')=1.25\times (r-K)-0.9,
\end{equation}
which is indicated as the solid line in Figure~\ref{nuvr}. 

This cut is approximately equivalent to a cut at a constant star formation 
rate in the total galaxy sample \citep[see e.g.,][]{W09UVJ} the black
$<\mathrm{sSFR}>$ arrow in the figure denotes the direction of increasing sSFR), 
and it is therefore quite effective at selecting galaxies with early-type SEDs. 
This method represents an alternative approach to the one suggested by 
\cite{AWLF07} and is a similarly powerful tool to identify dusty 
galaxies among the early-type galaxy population.

\subsection{Completeness and Contamination}\label{ccl}

\begin{figure}
 \centering
 \includegraphics[width=0.5\textwidth]{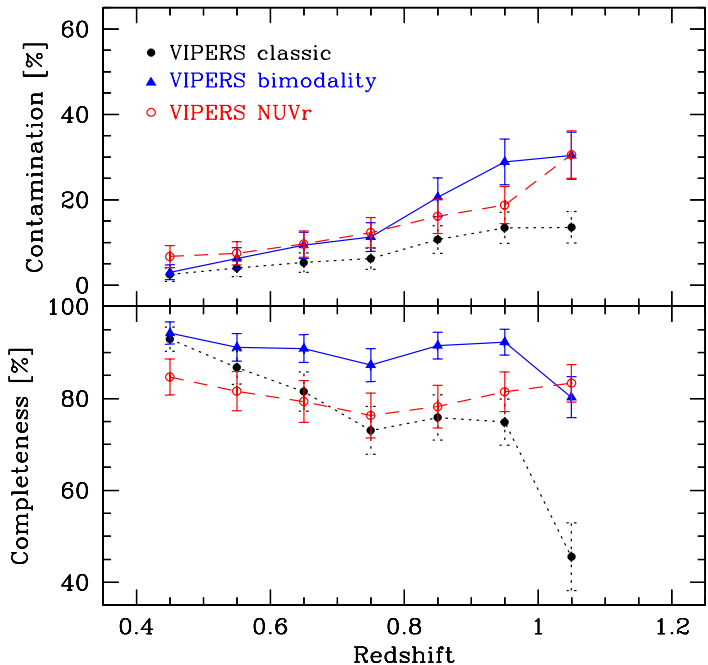}
\caption{Completeness of early-type galaxies (lower panel) and contamination 
by late-type galaxies (top panel) with respect to the SED type-classified
early-type galaxy population as a function of redshift for different 
selection methods. Classical approach (dotted line, filled circles), 
colour-bimodality (solid line, blue triangles), $NUVr'$ sample
(dashed line, open circles).}
\label{cont}
\end{figure}

In the following analysis, we adopt the SED type classification as our 
reference selection method (see Section~\ref{type}).
For each of the other selection criteria we derive their completeness in
early-type galaxies (i.e., the fraction of the SED early-type galaxies which are
classified as early-type galaxies with that particular selection criterion) and 
the contamination due to late-type galaxies (i.e., the fraction of galaxies 
classified as early-type with that particular selection criterion which are
instead classified as late-type galaxies on the basis of the SED type
classification). Figure~\ref{cont} illustrates the early-type
completeness and late-type contamination as a function of redshift for each
selection criterion. Different lines give the completeness of red galaxies and
contamination of non-red galaxies for the various selection methods,
classical approach (dotted line and filled circles), colour-bimodality 
(solid line and blue triangles), $NUVr'$ selection (dashed line and 
open circles). 

Our colour-bimodality selection of early-types proves to be highly efficient
($\sim$85\%), even  up to the highest redshifts. Moreover, the contamination
by late-types is $<$10\% up to $z=0.8$ and reaches $\sim$30\% in the highest
redshift bins. The classical approach produces a smaller contamination 
($\sim$5-20\%), however at the costly price of a much smaller completeness
($<$80\% at $z>0.7$). In the highest redshift bin ($z>1$) its completeness 
is only $\sim$45\%. In terms of completeness in early-types at high redshift,
the $NUV-r'$ selection is better than the classical approach, whereas at 
the same time keeping the contamination from late-type galaxies at a 
low level ($\leq$20\% up to $z=1.1$). For the highest redshift bin, the
contamination is larger and similar to that of the colour bimodality approach.
Across all redshifts, the fraction of early-types in the $NUV-r'$ method remains
quite stable at $\sim$80-85\%, even up to $z=1.3$. The colour bimodality
criterion has an almost constant completeness in early-type galaxies of
$\sim$90\% up to $z=1$, but it contains a slightly higher late-type
contamination of $\sim$5-10\% than the $NUV-r'$ sample.

%

\section{The Evolution of the CMR since $z\sim1.3$}\label{ev}

\subsection{Fitting the Red Sequence}\label{fits}

For galaxies undergoing a pure passive evolution of their stellar populations
it is quite simple to predict their photometric evolution in terms of 
luminosity and rest-frame colours. By comparing these predictions with the
observed evolution of galaxies on the RS, it is therefore possible to place
some constraints on the evolutionary path of the RS objects. One method often
adopted to quantify the evolution in the observed properties of these
galaxies is to consider the change in the best fitting linear relation between 
luminosity and rest-frame colour, i.e. the evolution of the average colour of 
the RS objects at a fixed absolute magnitude 
\citep[e.g.,][]{Bel04b,Wei05,Fra07}. The apparent simplicity of this 
analysis is however hiding important sources of uncertainty in the 
interpretation of the observed evolution. The most important one is of course
the operational definition of the RS galaxies, as discussed in the previous 
sections, which in turn results into an uncertainty on the contaminating 
fraction of star-forming galaxies. We also need to consider differences in 
the magnitudes used to derive the observed and the rest-frame colours 
(aperture magnitudes versus pseudo-total ones), and also the priors used to 
constrain the fit. Finally, many data sets at high redshift suffer from low
sample statistics that do not allow for a reliable determination of the slope
of the RS relation, and therefore the fit is carried out using a fixed slope
at all redshifts.

This approach was adopted in the case of the COMBO-17 photometric galaxy survey 
by \cite{Bel04b}, who established the following evolution of the RS relation:
\begin{equation}
(U_J-V_J)_{\rm RS}=1.40-0.31z-0.08\ (M_V-5\ {\rm log}\ h+20.0),
\label{B04}
\end{equation}
where $h=0.7$, assuming a fixed slope of $d(U_J-V_J)_{\rm RS}/d(M_V)=-0.08$,
as derived for elliptical galaxies in the local Coma cluster (BLE92).
Another variation of this definition is a fixed (non-evolving)
parallel cut on the blue part of the CMR that globally separates passive red
(non-SF) galaxies from blue SF galaxies \citep[e.g.,][]{WAB09,ECDFSRS,Bra09,Whi10}.

%
%
\begin{figure}[t]
 \centering
 \includegraphics[width=0.5\textwidth]{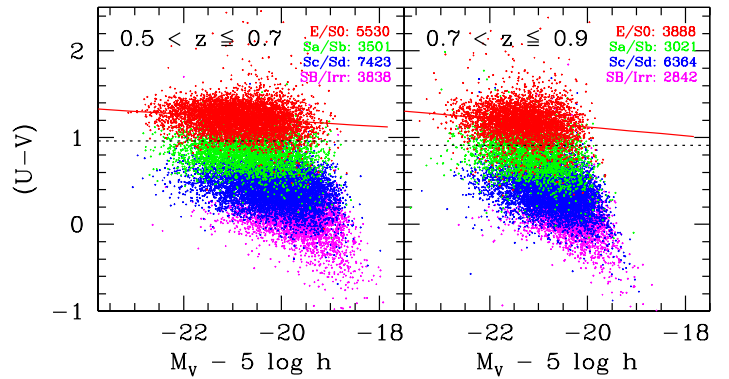}
 \caption{Evolution of the RS using a variable slope with redshift.
 The CMR is divided into two redshift bins at low 
 ($0.5<z\leq0.7$) and high ($0.7<z\leq0.9$) redshift. Galaxies are 
 colour-coded with respect to their derived SED type. The number of galaxies
 for a specific SED type is shown in each slice. The solid line is 
 the best-fit to the RS adopting the slope as derived from VIPERS within
 the selection criteria indicated with the dotted line (see text for details).}
 \label{CMRv}
\end{figure}

Local E+S0 galaxies in field and group environments display both a similar 
slope ($d(U_J-V_J)_{\rm RS}/d(M_V)=-0.05$) and scatter 
($\delta(U_J-V_J)=0.06$ mag \citep{SS1992} across the CMR as
their counterparts in clusters. However, it is still unclear
if the assumption of a fixed slope holds also for field galaxies at
higher redshift.

If we adopt the same methodology to fit the RS sample selected using
the classical approach described in section~\ref{class}, we obtain a fit to
the RS which is illustrated by the dotted red lines shown in the panels of
Figure~\ref{cmra}. It is evident that this best-fitting relation with a fixed
Coma cluster slope is somewhat too steep for the VIPERS data. We should consider
that our colours are based on the CFHTLS \texttt{mag\_auto} magnitudes 
that were measured in Kron-like elliptical apertures (see Section~\ref{phot}), 
and which are an approximation for total 
magnitudes\footnote{\cite{BA96} suggest that Kron-like aperture 
measurements miss a somewhat uniform $\sim$6\% of the
total flux for galaxies.}, while those used by BLE92 were based on fixed-aperture 
photometry. A difference in slope is therefore somewhat to be expected, as 
demonstrated by \cite{Sco01}: a CMR constructed using colour measurements
within fixed apertures is steeper than a CMR based on colours established 
with apertures covering the effective radius of the galaxy 
(or pseudo-total magnitudes, for that matter). 

Since the combination of large size and accurate photometry of VIPERS 
allows us to derive with high precision the slope and scatter of the RS
as a function of redshift, we decided to explore the different
results one would obtain with and without the fixed-slope constraint on the
fitting of the RS. Figure~\ref{CMRv} illustrates the CMR for VIPERS for
two redshift slices at $0.5<z\leq0.7$ and $0.7<z\leq0.9$. When we define RS 
galaxies using the bimodality in colour distribution, as described in 
Section~\ref{green} (the partitioning colour is identified in 
Figure~\ref{CMRv} by the dotted line), we obtain the following RS best-fitting
relations at redshift $\langle z\rangle=0.55$:
\begin{equation}
(U-V)_{\langle z\rangle=0.55}=(0.669\pm0.063)-(0.026\pm0.003)\times (M_V-5\ {\rm log}\ h),
\end{equation}
and at redshift $\langle z\rangle=0.80$:
\begin{equation}
(U-V)_{\langle z\rangle=0.80}=(0.545\pm0.029)-(0.029\pm0.005)\times (M_V-5\ {\rm log}\ h).
\end{equation}
For consistency with the classical method, we fitted a free variable
slope to the whole redshift interval at $0.4<z\leq1.3$ probed by VIPERS, giving
similar results within the uncertainties as shown for the two redshift slices
in Figure~\ref{CMRv}. We therefore conclude that there is no evidence of
an evolving slope from $z=1$ to $z=0.4$. 

The slope we obtain is significantly flatter than the one obtained for the 
Coma cluster by BLE92, but entirely compatible with the flatter slope 
obtained by \cite{Sco01} using variable apertures to cover a fixed fraction 
of the galaxy total light. However, the effect that the different slope might
have on the observed evolution of the properties of the RS turns out to 
be almost negligible. The mean RS colour that is deduced from the 
best-fitting linear relation where the bulk of the RS galaxies are located
(absolute magnitude range between $-22$ and $-20$), changes by less than 
0.05 magnitudes, which is significantly less than the redshift evolution
discussed in the following section.

%
%
\begin{figure}[t]
 \centering
 \includegraphics[width=0.5\textwidth]{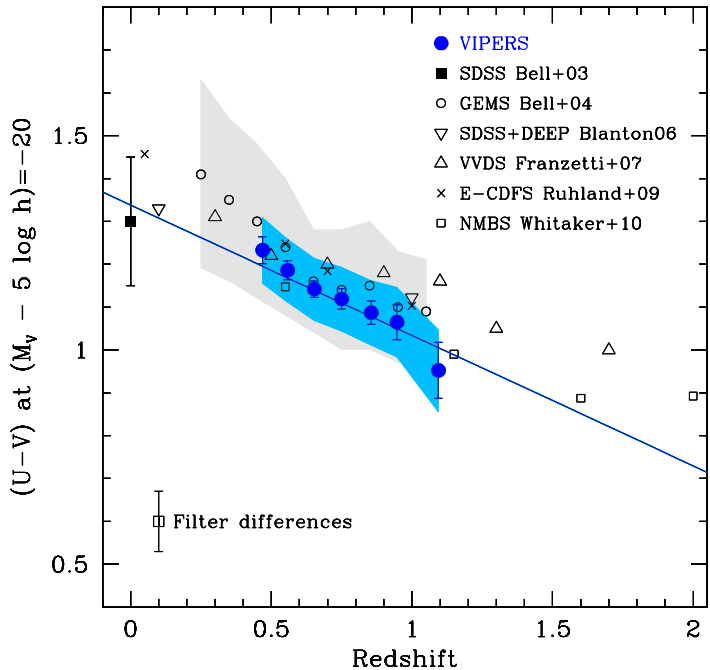}
  \caption{Evolution of the CMR intercepts in $(U-V)$ at
  $M_V-5\ {\rm log}\ h=-20$ for red-sequence galaxies as a function of
  redshift. The solid line is a least-square fit to the VIPERS data including
  the local SDSS reference (filled black square).
  The blue area reflects the total observed scatter due to systematic 
  uncertainties, whereas uncertainties due to Poisson noise are indicated with
  single error bars. The grey shaded area shows the observed measurement errors 
  of the COMBO-17 data set by \cite{Bel04b}. In the bottom left typical
  uncertainties due to filter and rest-frame transformations are shown.}
 \label{CMRev}
\end{figure}

\subsection{Evolution of the CMR intercept}\label{int}

Figure~\ref{CMRev} displays the evolution of the RS for the VIPERS sample
(filled circles) as a function of redshift. To allow a fair comparison
of our results with previous works, the evolution of the RS has
been computed following the classical RS approach (see Section~\ref{class} and
\ref{fits}). The literature data comprise various surveys,
using either spectroscopic ($z_{\rm sp}$) or photometric redshifts
($z_{\rm ph}$): the COMBO-17 sample \citep{Bel04b}, $z_{\rm ph}$, 
open circles), a study in the Extended Chandra Deep
Field South (E-CDFS, \citep{ECDFSRS}, $z_{\rm sp}$, crosses),
an SDSS and DEEP2 analysis \citep{Bla06}, $z_{\rm sp}$, inverted triangles),
the multi-wavelength medium NIR-band photometric study of 
NEWFIRM Medium-Band-Survey (NMBS, \citep{Whi10}, $z_{\rm ph}$,
open squares), and the results of the VVDS \citep{Fra07},
$z_{\rm sp}$, triangles). The solid line is a fit to the VIPERS data 
but also including the SDSS measurement \citep{Bel04b}, filled black
square) as a local reference point at $z=0$. Note that \cite{Bla06} derived a 
slightly different measurement for SDSS (inverted triangle) which is consistent
with the SDSS scatter. The measurement uncertainties on the RS intercepts which
arise from Poissonian statistics were derived using bisector fits with the 
errors on the bisectors being evaluated through a bootstrap  resampling of the 
data \citep{FZBSD05}. The total observed measurement errors
including systematic uncertainties due to filter transformations
and zero-point variations in the photometry are shown with the blue
error corridor. For comparison purposes, the observed measurement error
bars of the COMBO-17 data \citep{Bel04b} are  shown as a grey 
shaded area.

\begin{table}[t]
\caption{$(U-V)$ colour intercept evolution at $M_V-5\ {\rm log}\ h=-20$
as a function of redshift for a fixed slope of 
$d(U-V)/d(M_V)=-0.08$. Columns give 
the median redshift, number of galaxies used for the computation,
intercept, measured scatter (Poissonian statistics), observed scatter
(measured plus systematic scatter), and the observed colour dispersion of
the CMR $\sigma (U-V)$ between $-19.0\ge M_V\ge -23.5$.}
\label{RSint}
\centering                         
\begin{tabular}{c c c c c c}   
\hline\hline
$\langle z\rangle$ & $N_{\rm gal}$ & Intercept & Measured & Observed & Colour \\
 & & & Scatter & Scatter & Dispersion \\ 
\hline
0.47 & 1292 & 1.23 & 0.032 & 0.077 & 0.143 \\
0.56 & 2661 & 1.19 & 0.022 & 0.074 & 0.135 \\
0.65 & 3360 & 1.14 & 0.019 & 0.073 & 0.128 \\
0.75 & 2420 & 1.12 & 0.024 & 0.075 & 0.148 \\
0.86 & 1583 & 1.09 & 0.027 & 0.076 & 0.161 \\
0.95 &  771 & 1.07 & 0.041 & 0.082 & 0.135 \\
1.09 &  250 & 0.95 & 0.066 & 0.097 & 0.100 \\
\hline                        
\end{tabular}
\end{table}

The average evolution of the VIPERS RS for the classical selection
criterion can be expressed as
\begin{equation}
(U-V)=1.34\pm0.03-0.30\pm0.04 \times\ z -0.08\  (M_V-5\ {\rm log}\ h+20.0).
\end{equation}
Table~\ref{RSint} lists the $(U-V)$ colour intercept evolution at 
$M_V-5\ {\rm log}\ h=-20$ as a function of redshift for a 
fixed slope of $d(U-V)/d(M_V)=-0.08$.
If the VIPERS sample is split into sub-samples for the W1 and W4 fields,
a similar evolution of the RS with similar uncertainties is found.
Uncertainties due to cosmic variance have only a weak influence on
the derived evolution.

Overall, the VIPERS results themselves show a similar evolution with redshift
compared to various surveys in the literature which adopted the classical 
approach but with a significantly higher precision and statistics.
Galaxies located on the RS display a moderate evolution of
the intercept of $\Delta (U-V)=0.44\pm0.12$ since $z\sim1.3$ up to the
present-day. Across the redshift interval $0.4<z<1.3$ the average
RS intercept of VIPERS evolves as $\Delta (U-V)=0.28\pm0.14$.
This result alone is inconsistent with stellar population synthesis predictions
of $\Delta (U-V)=0.09\pm0.02$ that assume a high formation redshift for the
bulk of the stars at $z_{\rm form}=4$ and a subsequent passive evolution of 
their stellar populations across the redshift range $0.4<z<1.3$ 
(see Section~\ref{sp}). However, there could be a small systematic 
difference in the adopted photometric zero-point (most likely due to the 
different $U$-band filters) which partially contributes to the intercept 
evolution at $z\gtrsim0.8$. The most significant difference visible in 
Figure~\ref{CMRev} is at redshift $z\sim 1.1$ between the VIPERS and VVDS
point, but differences in sample selection 
and area coverage (the VVDS being a purely magnitude
limited survey to $i'_{\rm AB}<24.0$ over $\sim$0.7 deg$^2$),
can totally explain this difference.
At lower redshift the VIPERS data are entirely consistent with previous 
measurements. 

We have tested the impact of various systematic effects on our data.
In general, possible systematic differences in the evolution of
the intercept among the different surveys are most likely either due to sample
statistics, differences in the filter systems or the impact of cosmic variance.
In Appendices~\ref{compl} and \ref{cvar} we discuss the completeness 
of the VIPERS sample at high redshift and the systematic uncertainties
introduced due to cosmic variance.
VIPERS comprises the largest number of galaxies between $0.5<z<1.2$ 
and at the same time covers the widest sky area in comparison to
previous galaxy surveys. Therefore, both restrictions due to low number 
statistics or cosmic variance can be largely ruled out. Nevertheless, 
we have estimated the systematic effect of cosmic variance using the 
prescription by \cite[][see Appendix~\ref{cvar} for details]{MSNR11}.
For RS galaxies between $0.4<z<1.3$, the uncertainties arising from cosmic 
variance vary in the range $0.04<\log(M_{\star}/M_{\odot})<0.07$, with a median
of $\langle\log(M_{\star}/M_{\odot})\rangle=0.05$.
We have also tested the impact on our findings of practical choice, e.g.,
using different redshift bin sizes, but our choice does not 
affect the results within the measured uncertainties. A quantitative 
comparison with previous works is complicated by the fact that many surveys do 
not provide the exact prescription of their used filter transmission curves.
This introduces uncertainties in the adopted filter passband 
transformations and conversion to rest-frame magnitudes when comparing
different surveys, which depend on their used spectral templates 
(see also Appendix~\ref{uband}). At the left bottom of Figure~\ref{CMRev} 
the typical average uncertainty in filter transformations is indicated with 
an error bar. This uncertainty combines contributions from using different 
$U$ and $V$-band filter passbands and uncertainties introduced from the
conversion to absolute rest-frame colours.

\begin{figure}[t]
 \centering
 \includegraphics[width=0.5\textwidth]{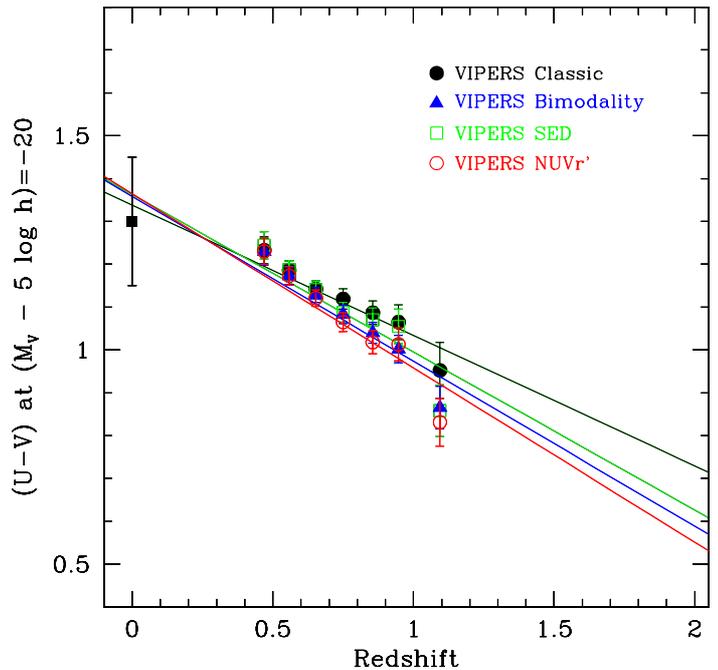}
  \caption{Evolution of the RS for different selection criteria of early-type
  galaxies. The colour evolution of the red-sequence as described by the
  CMR intercepts in $(U-V)$ at $M_V-5\ {\rm log}\ h=-20$ as a function of redshift.
  The evolution of the RS is shown for different selection criteria
  colour-bimodality (blue filled triangles), SED type classification (red
  open squares) and the classical method (black filled circles). 
  Measurement error bars are displayed for each criteria. All relations are
  least-square fits to the respective VIPERS selection criteria including the
  local SDSS reference (filled black square).}
 \label{CMRm}
\end{figure}

In Figure~\ref{CMRm} we compare the measured evolution of the RS with redshift
for different selection criteria, but keeping a fixed, non-evolving intercept
as for the classical method. The RS evolution is shown for the colour-bimodality
criterion (blue filled triangles), SED type classification (green open squares),
the $NUVr'$ selection (red open circles), and the classical method
(black filled circles). In comparison to the classical approach, 
all other methods show a steeper evolution. It is somewhat expected that 
the classical approach displays a weaker evolution as this method is biased
towards the few most brightest and reddest galaxies, whereas the other
criteria consider the overall early-type galaxy population including
also early-type galaxies with bluer colours. It is interesting to note 
that among the alternative selection criteria of early-type galaxies
there are not large variations. The method based on the SED types
produces results very similar to those obtained by the colour-bimodality and 
the $NUVr'$ selection. The good agreement among these different
selection methods suggests that our early-type galaxy sample contains
little amount of dust and that the effects of dust obscuration become more 
dominant at higher redshifts ($z>1.3$). 

Apart from different selection criteria, internal colour variations in
the galaxies might impact the derived results. For bright, low-redshift 
galaxies ($z\lesssim0.45$), the adopted aperture in the photometry might 
restrict our colour measurements to the central (redder) parts of the galaxies.
Internal colour gradients could change the evolution of the RS intercept at 
the 10\% level, resulting in bluer average colours for low redshift galaxies
\citep[e.g.,][]{PDI90}. As throughout our analysis the same aperture 
size is adopted in all photometric bands, possible internal colour gradients
are neglected, which might give slightly steeper colour relations for
brighter galaxies. In particular, the effect of internal colour gradients might
cause our lowest redshift bin to display a slightly redder evolution of the
RS intercept compared to the redshift evolution found for the higher redshift
data.

%
%

%
%
\begin{figure}[t]
 \centering
 \includegraphics[width=0.5\textwidth]{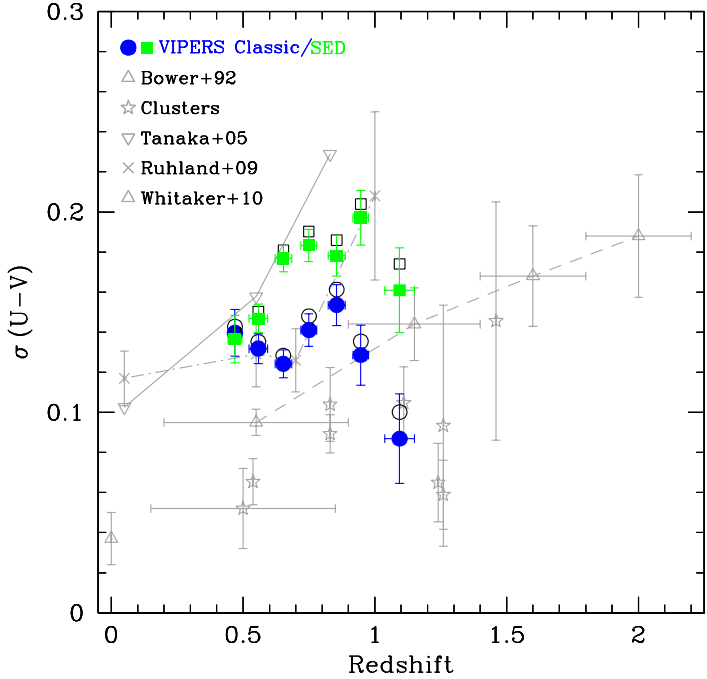}
 \caption{Intrinsic (filled symbols) and observed scatter (open symbols)
 in the rest-frame $(U-V)$ colours of passive galaxies.
 The scatter due to measurement uncertainties is subtracted from the
 observed scatter in quadrature. Literature data are shown in grey.
 Associated data for field galaxy studies (triangles or crosses) are joined 
 with lines, studies based on galaxy clusters are shown as stars.}
 \label{uvsca}
\end{figure}

\subsection{Scatter across the CMR}\label{sca}

Over the past years observations have found evidence for a rapid increase of the 
total stellar mass of passive red-sequence galaxies since $z\sim1$
\citep[e.g.,][]{Bel04b,Bro07,Bun05,Fab07,Sca07}.
This evolutionary process reveals 
itself in terms of an increase in the number density of quiescent 
sub-$L^*$ galaxies \citep[e.g.,][]{Bor06,Fab07}. 
One possible scenario is that the
red-sequence population is continuously supplied with a population of massive
blue galaxies that have their star-formation activity truncated or quenched over a
short time scale \citep{Bel04b,Bel07,Fab07}. Such processes
would manifest themselves primarily in a larger scatter in the CMR because the
continuous agglomeration of blue galaxies onto the RS will cause 
a broadening of the RS.

Figure~\ref{uvsca} illustrates the evolution of the scatter of the RS galaxies
as a function of redshift. We compute the intrinsic scatter (filled symbols) 
by subtracting the scatter due to measurement uncertainties from the
observed scatter (open symbols) in quadrature. The results are shown for the 
classical approach (blue circles) and the SED type selection (green squares).
Several interesting trends can be seen.
The scatter derived from the classical method is constant up to $z=1$, whereas
the scatter measured on the basis of the SED type classification is increasing
with redshift. Our highest-redshift measurement at $z=1.07$ is partly affected
by incompleteness effects and the real scatter is most likely larger by 
$\Delta (U-V)=0.05$ mag. The VIPERS sample is in good agreement with 
previous field galaxy surveys in the E-CDFS \citep{ECDFSRS} and 
the NMBS \citep{Whi10} that use a similar computation for 
the scatter of the RS, whereas field galaxies from the PISCES project
\citep{Tan05} display somewhat a larger scatter.
Interestingly, all field galaxy surveys predict a larger
scatter of the RS than the one obtained for cluster galaxies, both in the 
local galaxy cluster represented by the Coma cluster (open triangle, BLE92) 
and at higher redshift for several high-redshift galaxy clusters (star symbols).
The cluster measurements comprise data for 13 clusters in the literature
over the redshift range $0.18<z\lesssim1.46$ 
\citep{ESD97,Bla03,Bla06,Mei06a,Mei06b,vDFK08,Hil09}.
Despite possible differences in the colour measurements between cluster
and field studies (the former usually adopt fixed apertures and restrict the RS to
a few of the most luminous cluster members), the scatter in the rest-frame 
$(U-V)$ colour for quiescent galaxies in the field increases with redshift
and is a factor of more than two larger than in clusters at the same redshift.
This points towards a dependence of the scatter and its evolution on the 
environment. A possibility is that field galaxies consist of younger stellar 
populations compared to their cluster counterparts
\citep{BRCW98,FBZ09}. Alternatively, the larger
scatter in the field environment could be the result of a mixed population of 
old and young quiescent galaxies, whereas in clusters the RS is only populated
by older red galaxies that formed their bulk of stars at higher redshift.

%
%
\begin{figure*}
 \centering
 \includegraphics[width=0.8\textwidth]{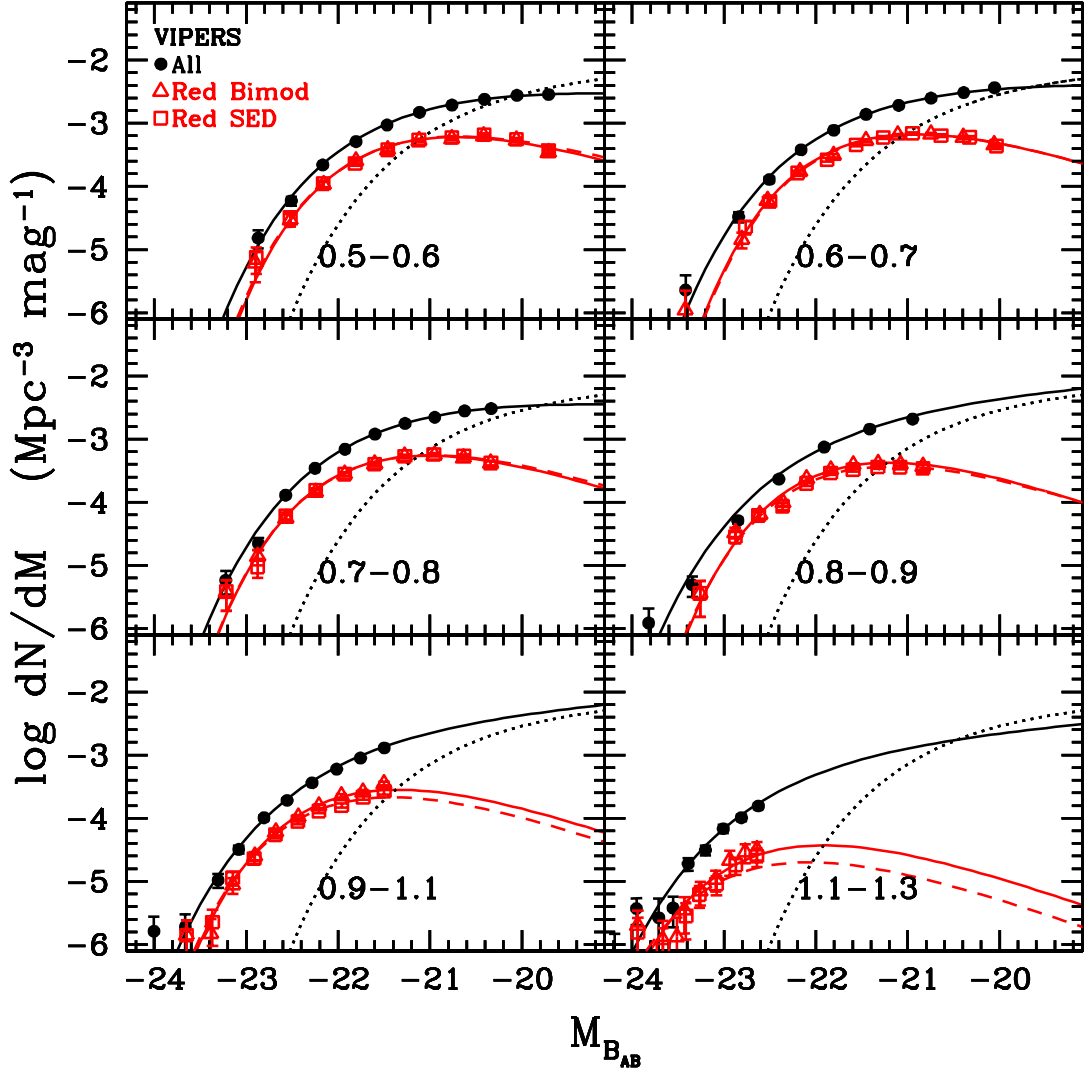}
 \caption{Rest-frame $B$-band luminosity function VIPERS
 from $z=0.5$ to $z=1.3$. The LF is shown for the total sample (black circles),
 for red galaxies defined using the colour-bimodality (red triangles, solid 
 line), and for red galaxies selected using the SED criterion (red squares,
 dashed line). Lines indicate the fits to the luminosity function using a
 Schechter function. For reference, the dotted black line shows the Schechter
 function of a local sample by \citep{ITZ05}.}
 \label{LFall}
\end{figure*}

\section{The Luminosity Function in VIPERS}\label{LF}

The luminosity function (LF) of field galaxies represents a fundamental
diagnostic to probe the star formation history of (blue) galaxies and the
gravitational mass assembly history of (red) passive galaxies. In particular,
since it has been speculated that the mass assembly of red early-type
galaxies could take place on different time-scales than the much earlier
formation of their  stellar populations \citep{BCF96,DF01}, 
the observed stellar mass-to-light ratios and the evolution of the CMR of passive 
galaxies could be non-reliable tracers of the mass assembly process, while 
the LF of RS galaxies would provide a much more sensitive diagnostic. 

In this work the LFs were computed using the ``Algorithm for Luminosity 
Function'' (ALF) as described in \cite{ITZ05}. ALF was originally 
developed for the VVDS project \citep{ITA04,ITZ05} and is a tool 
that implements several standard estimators of the LF: the non-parametric 
$1/V_{\max}$ \citep{Vmax68}, C+ \citep{Cplus71,Zuc1997}, and
SWML \citep{SWML88}, and the parametric STY \citep{STY79}. 
For the parametric STY estimator the empirical Schechter function
\citep{SF76} was adopted:
\begin{equation}
\phi(L)dL=\phi^*e^{-\frac{L}{L^*}}\left(\frac{L}{L^*}\right)^{\alpha}d\left(\frac{L}{L^*}\right).
\label{schechter}
\end{equation}
While the normalisation is directly done for the 1/V$_{\rm max}$ 
estimator, the SWML, STY and C$^+$ estimators  are independent of the spatial
density distribution and lose their normalisation. For these three estimators
we adopt the \cite{SWML88} density estimator to recover their 
normalisation \citep[see][for details]{ITZ05}.
The advantage of ALF is its capability to use the same data to provide
simultaneously the four different estimators, and this allows for an easy
verification of the robustness of the luminosity estimates against the effects
of binning (in absolute magnitude), of spatial inhomogeneities, or of incompleteness
in spectral type.
Absolute magnitudes were derived as discussed in Section~\ref{rest}.
For a given redshift, different galaxy types appear in different absolute
magnitude ranges because of the combination of the fixed apparent magnitude
limit used for sample selection with the $k$-corrections.
To avoid in particular a bias at the faint-end of 
the LF, the estimates of the LF are restricted to only
those galaxies within the absolute magnitude range for which the 
estimators are in agreement \citep{ITA04}. At magnitudes fainter than
this limit, the $1/V_{\max}$ estimator provides number densities which are
underestimated, therefore giving a lower limit of the LF slope.
The $\phi^*$ values are computed for each point of the $\alpha-M^*$ 
error contour at 1$\sigma$ confidence level. The uncertainty on $\phi^*$ 
is derived from the maximum between Poissonian error and error derived 
from the error contour.

All LFs are constructed for the rest-frame Johnson $B$-band, which is most
appropriate for the median redshift of the VIPERS sample at $z\sim0.70$.
This choice is driven by the fact that the dependency of the 
absolute magnitudes on the models is reduced and any possible bias arising
from different spectral types are minimized \citep{ITZ05}.
Furthermore, as the $B$-band is a common filter choice, a comparison to
previous results in the literature is straightforward.
Incompleteness in the survey was corrected for through the 
weighting scheme described in Section~\ref{swei}.

%
%
\begin{figure*}
 \centering
 \includegraphics[width=0.48\textwidth]{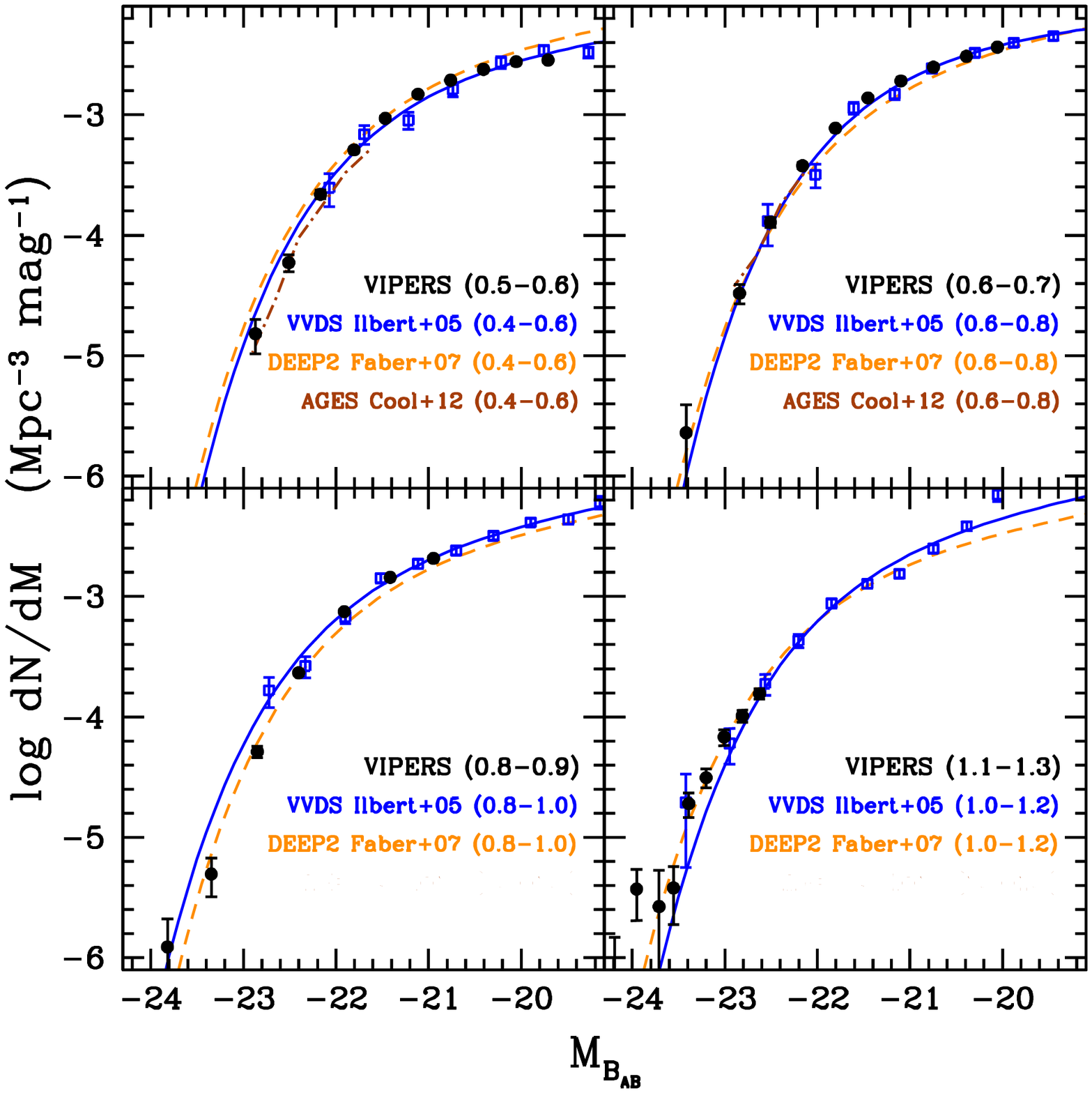}
 \includegraphics[width=0.48\textwidth]{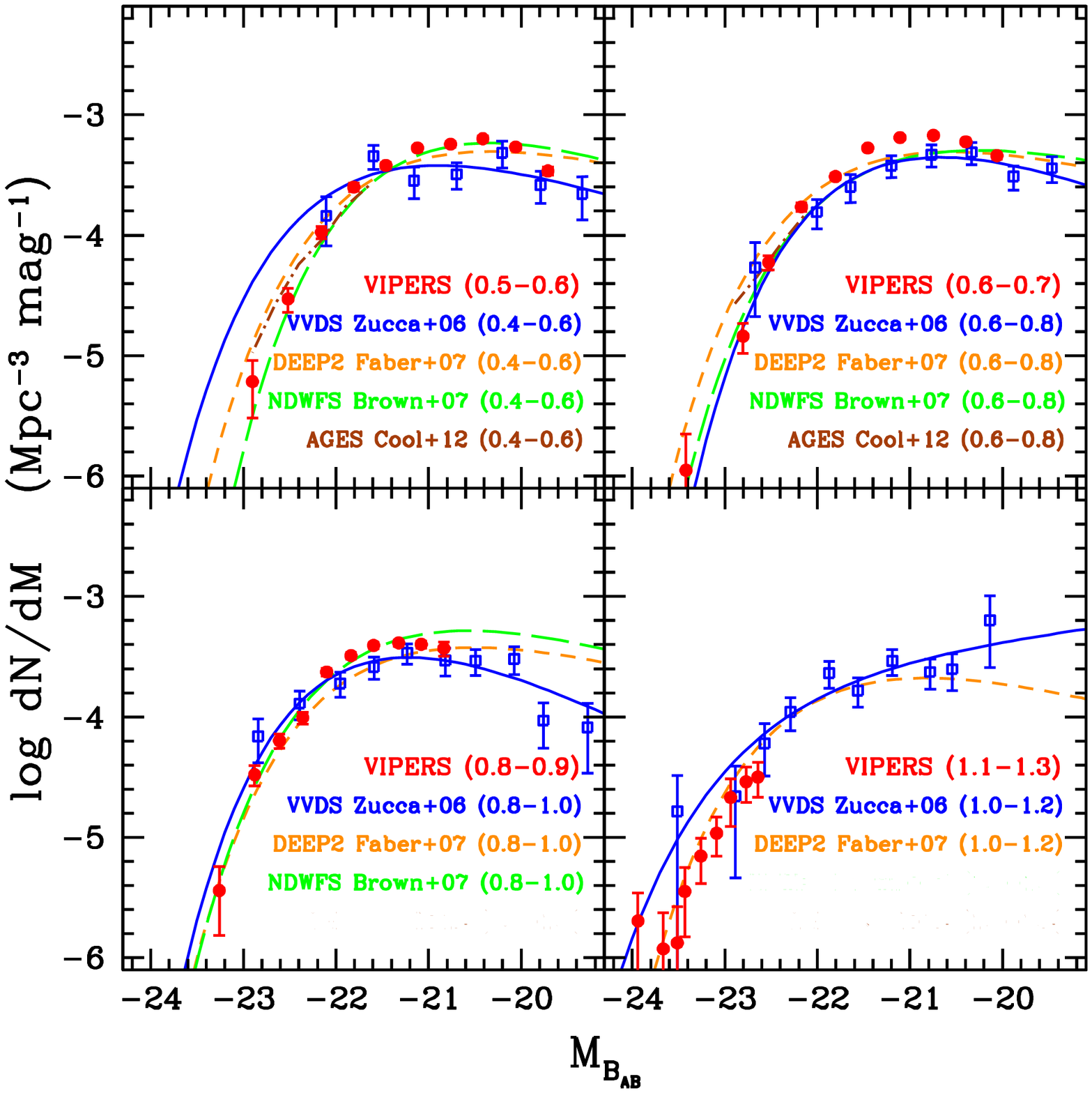}
 \caption{Comparison of the VIPERS $B$-band LF to previous measurements.
  Literature data include the LF of VVDS \citep[][blue squares]{ITZ05}, 
  the LF of DEEP2+COMBO-17 \citep[][orange dashed line]{Fab07}, 
  the LF of red early-type galaxies in the VVDS \citep[][blue squares]{ZIB06},
  the LF of red galaxies in the NDWFS \citep[][green long-dashed line]{Bro07}
  and the LF of galaxies in AGES \citep[][brown dot-dashed line]{CEK12}.
  The numbers in the brackets denote the redshift interval of the respective LF.
  {\it Left:} LF for all galaxies. {\it Right:} LF for red galaxies.} 
 \label{LFlit}
\end{figure*}
\begin{table*}[th]
\caption{Schechter parameters and associated one sigma errors 
($2\Delta\,{\rm ln}\,\mathcal L=1$) of the global VIPERS LF and the LF for different sub-samples
derived in the rest-frame standard $B$-band filter system. Parameters
listed without errors are set ``ad hoc'' to the given value.}
\label{LFs}
\centering                         
\begin{tabular}{c c c c c c}   
\hline\hline
\multicolumn{6}{c}{$\Omega_m=0.25$ \hspace{1cm} $\Omega_\Lambda=0.75$} \\ 
\hline
 & & & & & $\phi^{*}$ \\
Sample & $z$-range & $N_{\rm gal}$ & $\alpha$ & $M^*_{B}-5\log~(h)$ & $(10^{-3}~h^{3}~{\rm Mpc}^{-3})$ \\
 & & & & &  \\
\hline
All & 0.5-0.6 & 7580 & -0.78$\pm0.05$ & -19.98$\pm0.05$ & 16.21$^{+0.72}_{-0.73}$  \\
    & 0.6-0.7 &10386 & -0.90$\pm0.05$ & -20.20$\pm0.04$ & 17.70$^{+0.79}_{-0.80}$  \\
    & 0.7-0.8 & 9066 & -0.81$\pm0.06$ & -20.19$\pm0.05$ & 18.44$^{+0.77}_{-0.81}$  \\
    & 0.8-0.9 & 5973 & -1.29$\pm0.09$ & -20.61$\pm0.07$ & 11.93$^{+1.16}_{-1.18}$  \\
    & 0.9-1.1 & 4217 & -1.30  	      & -20.67$\pm0.02$ & 11.27$^{+0.17}_{-0.17}$  \\
    & 1.1-1.3 &  337 & -1.30  	      & -21.02$\pm0.08$ &  4.90$^{+0.27}_{-0.27}$  \\
\hline
Red & 0.5-0.6 & 2089 & 0.18$\pm0.09$ & -19.69$\pm0.06$ & 5.11$^{+0.11}_{-0.11}$  \\
    & 0.6-0.7 & 2823 & 0.25$\pm0.09$ & -19.80$\pm0.05$ & 5.56$^{+0.11}_{-0.11}$  \\
    & 0.7-0.8 & 2330 & 0.27$\pm0.12$ & -19.90$\pm0.06$ & 4.53$^{+0.10}_{-0.13}$  \\
    & 0.8-0.9 & 1518 & 0.35$\pm0.21$ & -20.01$\pm0.09$ & 3.45$^{+0.24}_{-0.30}$  \\
    & 0.9-1.1 & 1091 & 0.30	     & -20.21$\pm0.03$ & 2.32$^{+0.07}_{-0.07}$  \\
    & 1.1-1.3 &   53 & 0.30	     & -20.77$\pm0.14$ & 0.31$^{+0.05}_{-0.05}$  \\
\hline
Red SED & 0.5-0.6 & 2143 & 0.10$\pm0.09$ & -19.72$\pm0.06$ & 5.26$^{+0.11}_{-0.11}$  \\
        & 0.6-0.7 & 2796 & 0.25$\pm0.09$ & -19.79$\pm0.05$ & 5.50$^{+0.11}_{-0.11}$  \\
        & 0.7-0.8 & 2380 & 0.21$\pm0.12$ & -19.92$\pm0.06$ & 4.67$^{+0.09}_{-0.11}$  \\
        & 0.8-0.9 & 1337 & 0.25$\pm0.22$ & -20.07$\pm0.10$ & 3.10$^{+0.18}_{-0.24}$  \\
        & 0.9-1.1 &  868 & 0.30 	 & -20.27$\pm0.03$ & 1.76$^{+0.06}_{-0.06}$  \\
        & 1.1-1.3 &   45 & 0.30 	 & -20.93$\pm0.16$ & 0.17$^{+0.03}_{-0.03}$  \\
\hline
\end{tabular}
\end{table*}

The VIPERS sample allows us to explore the evolution of the LF over a
significant redshift range, but in particular for a large sample with 
homogeneous selection function over a wide area. The $B$-band LF of the global
VIPERS sample is shown in Figure~\ref{LFall}. To be consistent with our analysis
of the GSMF (D13), we adopt here the same redshift bins used in that work. 
The symbols with error bars show the results obtained with the $1/V_{\max}$ 
method. The solid lines are the result of the STY estimator
\citep{STY79}, which is independent from the adopted bin sizes, and
where the likelihood of the present galaxy sample is maximized by assuming 
that the underlying LF can be parameterized with a \cite{SF76}  function. 
As a reference, in all panels the dotted black line shows the Schechter fit
to a local sample from the VVDS at $0.05<z<0.20$ \citep{ITZ05}. This
reference sample is comparable within $\Delta M_B<0.05$~mag to the local
SDSS sample of \cite{Bel03}. 

The evolution of the bright part of the LFs with redshift becomes evident at 
high-redshift ($0.8<z<1.2$) with $1/V_{\max}$ number densities being higher
than the Schechter function (see also Figure~\ref{LFlit}). 

Given our large sample, and our luminosity coverage, in the redshift range 
$0.5<z<0.8$ the faint-end slope of the STY estimator was determined as part of
the fitting process. Because the sample at higher redshifts becomes 
incomplete at faint luminosities, the slope was kept fixed at the values of 
$\alpha=-1.3$ and $\alpha=0.3$ for the total and red sample, respectively. 
When the two VIPERS fields are compared, the shape
and slope of the LF are very similar. At high-redshift ($0.8<z<1.2$), there is
some evidence for an underabundance of red luminous galaxies in W4 compared
to W1, with differences consistent with the effects of cosmic variance or 
environmental effects.

We have shown that the evolution of the RS depends to some extent on the
selection criteria of red passive galaxies. It is interesting, therefore,
to look if similar effects are observed in the LF. To explore the LF for
different galaxy types, we separated red and blue galaxies on the basis of
the colour-bimodality and the SED type classification. The $B$-band LF 
of the RS galaxies for the two different selection criteria are shown in 
Figure~\ref{LFall}. Both selection criteria give rise to very similar LFs, 
with only some differences at $z\gtrsim0.9$. The slope of the LF is flatter
for red passive galaxies compared 
to the global LF, while across the whole redshift interval $z=0.5$ to $z=1.3$ 
the red galaxy population contributes significantly to the bright part of the global LF.
These findings are in good agreement with previous surveys covering smaller 
areas, based on either photometric or spectroscopic redshifts  
\citep{ITA04,ITZ05,Cim2006,ZIB06,Bro07,Fab07,ZBB09}. In Table~\ref{LFs} we give for each redshift
bin the Schechter parameters and the corresponding one sigma errors measured 
with the STY estimator. Results are shown for the global LF and for the two separate
LFs for the red galaxy samples.

\subsection{Comparison with the Literature}

Figure~\ref{LFlit} shows the $B$-band LF of the total VIPERS sample 
(left panel) and the LF of red galaxies (right panel) compared to LF estimates
from the literature. For display purposes, only the red galaxy sample selected
on the basis of the colour-bimodality is shown. The literature data include the
LF of VVDS \citep[][blue squares]{ITZ05}, DEEP2+COMBO-17 data 
\citep[][orange dashed line]{Fab07}, and the LFs of early-type galaxies in the 
VVDS \citep[][blue squares]{ZIB06}, red galaxies in the NDWFS
\citep[][green long-dashed line]{Bro07},  and the LF of galaxies 
in AGES \citep[][brown dot-dashed line]{CEK12}. As the different LFs were
not constructed in exactly identical redshift intervals, the numbers in the 
brackets denote the redshift range of the respective LF.
Overall, we find a good agreement between the different LFs across 
the redshift range $0.5<z<1.3$, with possibly some differences for bright 
galaxies. The unique large volume probed by VIPERS at high redshift leads
to a higher completeness for rare, very luminous galaxies 
($M_V-5\ {\rm log}\ h<-23$) at high redshift ($z=0.8-1.3$), extending by
more than one magnitude the effective sampling of the bright-end of the LF.
As demonstrated in Section~\ref{CMR}, these very luminous objects are also
massive systems up to masses of $\log(M_{\star}/M_{\odot})\leq 11.7$
at $z\sim1$.

%
%
\begin{figure}[t]
 \centering
\includegraphics[width=0.5\textwidth]{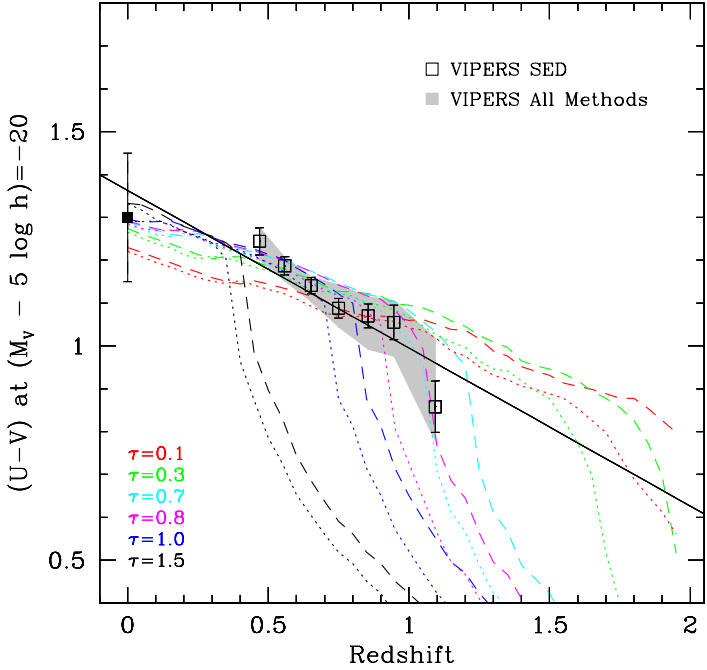}
  \caption{Comparison of RS intercept evolution with the 
 rest-frame colour evolution of stellar population models using different
 burst duration time scales $\tau$. Only PEGASE models with best-fitting models
 are shown. Dotted lines are for a $z_{\rm f}=2.5$, dashed lines for 
 $z_{\rm f}=3$. The shaded area denotes the variation of the observed
 evolution in the RS intercept for different selection criteria of 
 early-type galaxies.}
 \label{SPev}
\end{figure}

\section{Stellar Population Modelling of the Red-Sequence}\label{eev}

\subsection{Imprints on the evolution of the RS}\label{sp}

To assess the evolution of the RS as a function of redshift we
confront our measurements of the CMR intercept with the predictions of 
stellar population synthesis modelling.

In the context of single burst models, the evolution of the CMR intercept
follows the global trends of the expected passive reddening and fading of 
stellar populations formed at high redshift \citep{DF01,Fra07}. Variations 
of the CMR intercept as a function of 
redshift are mainly driven by the a priori assumed star formation history.
The intercept itself depends instead primarily on variations of the 
metallicity content of the stellar populations with minor contributions 
due to variations in the stellar population ages. 

We have constructed stellar population (SP) models using
the PEGASE2 library \citep{FRV97}.
The SP models are parameterised by a short-burst of SF starting at an initial 
formation redshift in the range $1.5<z_{\rm f}<5.0$ and with a duration 
$\tau$ in the range $0.1-3.0$ Gyr. The metallicity content in the models
is not held fixed but varies as the stellar population evolves with redshift.

Figure~\ref{SPev} shows the comparison of the predicted rest-frame
colour evolution for different stellar population models with our 
observational VIPERS results. For display purposes, only a limited range of models, 
those that best reproduce the observations, are shown. The dotted lines are for 
a formation redshift of 2.5, while the dashed ones are for a formation redshift of 3.
The VIPERS data rule out several model predictions with relatively high  confidence. 
Models where the most-recent star formation burst in galaxies was established 
at $z_{\rm f}\leq1.5$ predict too much evolution in the RS, and do not match our data.
Conversely, models with a formation redshift of $z_{\rm f}\geq4$
result in a much smaller evolution of the $(U-V)$ colour across the redshift 
range covered by our data, giving colours at $z=1$ that are
already too red and allowing only for a 
very modest colour evolution of $\Delta(U-V)\sim0.2$ mag to the present day.
Furthermore, the observations rule out SFHs with either a very extended burst ($\tau>0.8$), 
mainly because of the fact that the RS is already in place at $z\sim0.9$, or with
an extremely short one ($\tau<0.3$), mainly because in that case all the stars would have 
low metallicity and the resulting RS would be bluer than observed at $z<0.6$.
 Our data favour SFHs with $0.5\le\tau\le0.8$, although models with
$0.1\le\tau\le0.3$ and $z_{\rm f}=2.5$ cannot be completely ruled out.
These latter predictions also nicely describe the data at lower redshift
down to the present-day values indicated by local SDSS reference value 
(black filled square). In particular, the model prescriptions with 
$0.7\le\tau\le0.8$ precisely end up in the SDSS measurement at $z=0$.
From this comparison, we conclude that assuming a starting point at $z=1.3$,
the build-up of the RS happened rapidly within a very short time span of
only $\sim$1.5 Gyr. In such a scenario, early-type galaxies at high redshift
($0.9\lesssim z\lesssim1.1$) that have not yet moved to the RS
should still show signs of recent SF activity.

%
%
\begin{figure*}
 \centering
\includegraphics[width=0.8\textwidth]{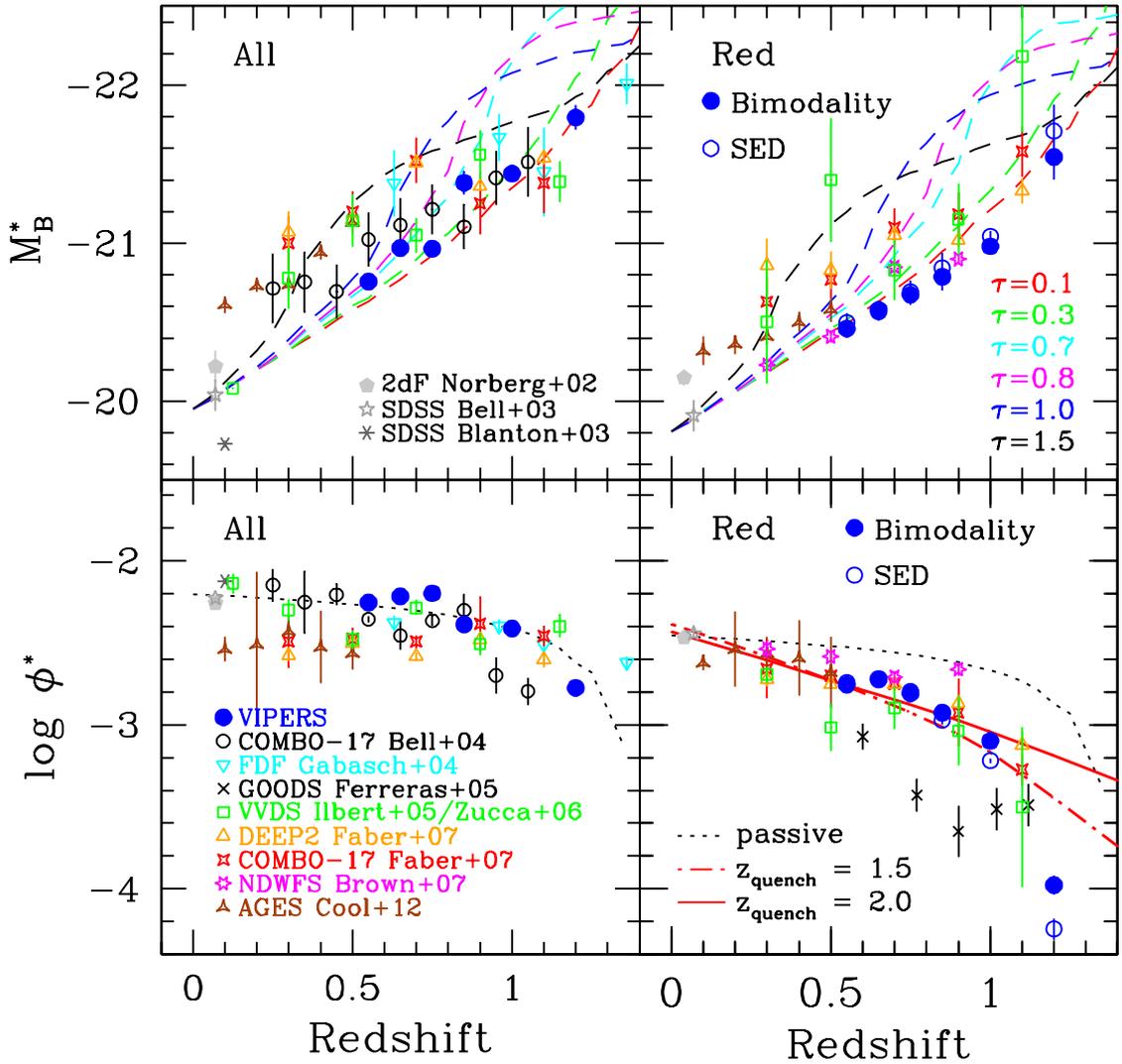}
 \caption{The evolution in $M_{B}^{*}$ (top panels) and in $\phi^{*}$
 (bottom panels) as a function of redshift for the VIPERS sample 
 (left), and for red galaxies only (right).  Various literature 
 data are shown for comparison. For the red LF, galaxies
 were divided into red and blue as defined using the colour-bimodality scheme
 (filled symbols) and according to their SED type classification (open symbols).
 In the top panels, the dashed lines indicate the evolution in $M_{B}^{*}$ 
 as predicted by the PEGASE models, calibrated on local SDSS data,
 with various burst duration time scales $\tau$. 
 In the bottom right panel the red lines give the predictions of quenching
 models with quenching formation times of $z_{\rm quench}=1.5$ (dot-dashed) and 
 $z_{\rm quench}=2.0$ (solid), which are in very good agreement with the VIPERS data. 
 The dotted line in the bottom panels gives the PEGASE prediction of 
 a passive evolution model for the evolution of the number density.}
 \label{LFmp}
\end{figure*}

\subsection{The Evolution of the Luminosity Function since $z=1.3$}\label{LFm}

The evolution of the LF parameters provides important information on
the evolution of the galaxy population and the different transformation
processes the galaxies experience. Changes in the characteristic Schechter 
magnitude $M^{*}$ reflect the aging of the stellar populations in galaxies, 
while changes in $\phi^{*}$ give information on the galaxy number density,
allowing us to estimate the fraction of objects that undergo merging or 
fading processes.

Figure~\ref{LFmp} shows the evolution of the Schechter function parameters
$M_{B}^{*}$ and $\phi^{*}$ with redshift for the total VIPERS sample 
(left panels) and for red galaxies only (right panels). 
In the top panels, the dashed lines indicate the evolution in $M_{B}^{*}$ 
as predicted by the PEGASE models with formation redshift $z_{\rm f}=2$
and various burst duration $\tau$ (see Section~\ref{sp}), which
were calibrated to match the local SDSS measurements.
For comparison, we also show literature data including
the COMBO-17 \citep{Bel04b}, FDF \citep{GBSHSF04}, field early-type 
galaxies in GOODS \citep{Fer05}, VVDS \citep{ITZ05}, 
early-type galaxies from the VVDS \citep{ZIB06}, and DEEP2 and 
COMBO-17 data sets \citep{Fab07}, red galaxies in the NDWFS
\citep{Bro07}, and galaxies in AGES \citep{CEK12}. Magnitudes 
of the different surveys were transformed to the concordance cosmology and
converted to Johnson-Cousins $B$-band magnitudes in the Vega system using 
$M_J=M_{AB}+0.084$ mag. For the total
VIPERS galaxy population $M_{B}^{*}$ brightens by $\sim$$1.04\pm0.06$~mag over
the redshift range $z=0.5$ to $z=1.3$ (see upper left panel), while RS galaxies
show a brightening of $M_{B}^{*}$ of $1.09\pm0.10$~mag. There are slight
differences in the $M_{B}^{*}$ evolution between red galaxies classified with
the bimodality and the SED type classification. The SED type classification
is in better agreement with the PEGASE models as in this selection criterion
galaxies with somewhat younger ages are also considered. Overall, the number 
density $\phi^{*}$ for red galaxies rises over the past 9~Gyr, with a 
stronger increase from $z=1.3$ to $z\sim0.7$, followed by a weaker rise from
$z<0.7$ to the present-day. This result confirms our findings on the evolution
of the RS and indicates first a rise in the build-up of massive galaxies 
at $0.7\lesssim z\lesssim 1.3$ over a relatively short time period of $\sim$1.5~Gyr,
followed by a continuous slower assembly over cosmic time.

To probe in more detail the mass assembly of the RS, we compare in 
Figure~\ref{LFmp} the number density evolution of red, quiescent galaxies to
predictions from stellar population synthesis models that involve a shut down 
of star formation, such as a sharp truncation or a quenching process. 
Sharp truncation models, resulting in a purely passive evolution of the
stellar population, can reproduce quite well the evolution of $M^{*}$, but
they fail to explain the build-up of the number density. An example of such 
a model prediction, constructed with a simple stellar population with a 
single metallicity formed at $z=3$ is shown by the dotted line in the bottom
panels of the figure. In contrast, quenching models predict a gradual 
increase in the RS galaxies number density with time. Here we have taken the 
quenching models from \cite{Har06}, which consist of a 1~Gyr long burst 
of star formation at $z=5$, which consumes 80\%-99\% of the gas reservoir, 
followed by progressive quenching of the star forming galaxies starting at 
$z_{\rm quench}=1.5$ or $z_{\rm quench}=2$. We notice
that the resulting star formation history for these models is actually quite 
similar to the one we used to derive the models used in the previous section 
to describe the evolution of the RS colour. The $\phi^{*}$ of the quenched models
was normalised to the observed $\phi^{*}$ of red galaxies at $z=0.5$. Note that 
despite the simple assumptions in the models, they describe the observed trends
reasonably well.

The lower right panel of Figure~\ref{LFmp} shows that the evolution of the red galaxy
populations in VIPERS follows quite closely the tracks for quenching models with
quenching onset at $z_{\rm quench}=1.5$ (dot-dashed line), with possibly
an indication that an even later onset of the quenching could better represent
the highest redshift bin data. A similar picture can be drawn from other surveys
\citep{Fer05,ZIB06,Fab07}, although
these data show a larger scatter because of higher measurement uncertainties. 
For the VIPERS colour bimodality sample the $\phi^{*}$ of the RS galaxies
rises by a factor of $4.4\pm0.9$ from $z=1$ to the present-day, whereas for
the SED-sample the $\phi^{*}$ of RS galaxies increases by a factor of 
$5.8\pm0.9$ over the same time interval. The $\phi^{*}$ of the quenching
model rises by a factor of 3.8 for $z_{\rm quench}=2.0$ and by a factor of
5.1 for $z_{\rm quench}=1.5$. Both model values are within the observed ranges,
although the VIPERS data at high redshift ($z>0.9$) prefer the predictions
with $z_{\rm quench}=1.5$. We find therefore another indirect indication 
towards early-type galaxies at high redshift ($0.9\lesssim z\lesssim1.1$) 
still showing signs of recent SF activity.

%
%
\begin{figure}[t]
 \centering
\includegraphics[width=0.48\textwidth]{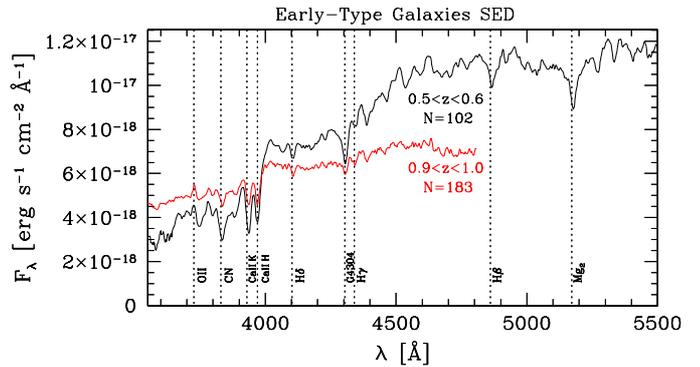}
 \caption{Example of stacked spectra of early-type galaxies at
 $0.5<z<0.6$ (black) and $0.9<z<1.0$ (red, shifted for display purposes).
 All spectra should represent passive, red and quiescent galaxies. 
 Early-type galaxies at high redshift show clear evidence of ongoing SF which 
 is supported by weak \oii\,3727 emission. Early-type galaxies have been 
 selected based on the SED type classification. Spectra of early-type 
 galaxies selected using other criteria look very similar. Prominent 
 absorption and emission features in the spectra are denoted with dotted lines.}
 \label{spec}
\end{figure}

\subsection{Evidence of enhanced Star Formation in the Spectral Properties of the Early-Type Galaxy Population}

To verify the indirect evidence for a recent star formation activity
among redshift $z\sim1$ RS galaxies, discussed in the previous sections,
we analyse directly the properties of these galaxies by stacking 
their spectra. Figure~\ref{spec} illustrates
examples of (rest frame) stacked early-type galaxy spectra at low-redshift 
($0.5<z<0.6$, black) and at high-redshift ($0.9<z<1.0$, red) selected on the 
basis of the SED type classification. Spectra of RS galaxies selected using 
the colour-bimodality method or the $NUVr'$ criteria are virtually indistinguishable 
from the ones shown. For the stacking process early-type galaxies within an absolute
magnitude range of $-22.5\ge M_V\ge -23.0$ were selected. The stacked spectra
comprise a similar number of individual spectra ($N_{\rm spec}\sim100$ at 
$0.5<z<0.6$, $N_{\rm spec}\sim180$ at $0.9<z<1.0$) and represent the typical
averaged  spectra of early-type galaxies at the respective redshift range.
At low-redshift the stacked spectrum is that of a purely passive, red and 
quiescent galaxy. However, the high-redshift counterpart displays clear
signs of ongoing SF as indicated by the detection of a weak \oii\,3727 emission
line. 

To quantify and constrain the recent star formation in 
the stacked spectra of early-type galaxies at high-redshift, 
the main questions of interest are the SFR, 
the mass fraction involved in the star formation and the duration 
of the star formation episode.
As the individual spectra used in the stacking have been re-normalized 
to reproduce the total flux observed photometrically for each galaxy,
we assume the measured \oii\,3727 equivalent width to be representative 
for the global value of the average VIPERS galaxy.
Transforming the observed \oii\,3727 equivalent widths to rest-frame and
assuming the relationship by \cite{Ken92b} with a global
extinction correction of E(H$\alpha$)=1~mag, we derive for the 
whole sample of early-type galaxies a median SFR of 
$1.8$ $M_{\odot}\,{\rm yr}^{-1}$. The different stacked spectra
result in similar SFRs estimates:
$NUVr'$:           median $\langle \mathrm{SFR}\rangle=1.9$ $M_{\odot}\,{\rm yr}^{-1}$, 
colour-bimodality:  $\langle \mathrm{SFR}\rangle=2.2$ $M_{\odot}\,{\rm yr}^{-1}$, 
SED types:	    $\langle \mathrm{SFR}\rangle=1.3$ $M_{\odot}\,{\rm yr}^{-1}$.
This suggests that a fraction between 1 and 3\% of the stellar mass of each 
galaxy is on average involved in the star formation burst which will exhaust 
the gas within $\sim$0.10-0.14~Gyr, adopting a burst duration time scale of
$t_{\rm d}\propto  M_{*}/\mathrm{SFR}$ \citep{NFW07}. 

In the nearby universe a non negligible fraction of passive galaxies 
on the red sequence shows signs of emission lines with characteristic 
line ratios of low ionisation nuclear emission-line regions
\citep[LINERs,][]{YFKKD06,Sch07}. Red galaxies
with strong LINER features exhibit on average younger ages than their passive
counterparts without emission, suggesting a connection between the local galaxy
activity and internal processes [e.g., shock ionization by starburst
winds, photoionization by (post-AGB) stars or AGNs] or external mechanisms
that could be responsible for the emission enhancement [e.g., cooling flows,
photoionization by low-luminosity-AGNs (LLAGNs), shock ionization due to 
gas accretion, galaxy interactions and merger activity]. 
Because of the spectral coverage, VIPERS galaxy spectra at high redshift
are restricted to the \oii\,3727 emission line as a star formation proxy. 
To quantify a possible contribution of LLAGNs in our red galaxy sample,
we have analysed the morphologies of massive galaxies with $M_{\odot}>11.2$.
The majority of these objects appear to be indeed isolated and massive systems. 
There is only a small fraction ($<20$\%) of galaxies which shows some signs of 
disturbed morphologies or possible galaxy interaction with a close neighbour.
This suggests that external mechanisms play only a minor rule in triggering 
star formation activity. It is outside the scope of the current work to
speculate about possible physical triggering mechnanisms for the quenching of
star formation. In the future, we will explore in more detail which processes 
are responsible and contribute to the star formation/AGN activity detected
in our galaxies.

Our findings support the scenario for the build-up of
the RS described in Section~\ref{sp}, where massive early-type galaxies show 
evidence for ongoing SF activity at $z\sim1$ and subsequently experience an
efficient shut-down mechanism of their SF over a short time scale of 
$\sim$1.5~Gyr with a rapid exhaustion of their gas reservoirs.


\section{Discussion}\label{dis}

Our results of Figure~\ref{LFmp} and Table~\ref{LFs} indicate a modest but
significant increase in the number density of luminous massive galaxies 
($M^{*}>10^{11}M_{\odot}$)  by a factor of 2.4 over the past 
$\sim$9~Gyr. This is in agreement with the factor of 1.8 increase between
$z=1$ and $z=0.6$ found at corresponding masses in our analysis of the
VIPERS stellar mass function (D13), with lower-mass systems 
increasing over the same redshift interval by a factor of 2.5. This trend 
is independent of the selection criteria used to define the early-type 
(quiescent) galaxy population, and is qualitatively consistent with
the overall observed trend in the evolution of the stellar mass density 
since $z\sim 3.5$ \citep[][and references therein]{ITZ05,IMCLF13}, while 
being significantly different from the predictions of a purely passive 
evolution model.

The shapes of the LF in Figure~\ref{LFall} and the MF of D13
show that this evolution is strongly dependent on the luminosity and
mass of galaxies: the higher the luminosity (mass), the  
smaller the evolution. The diverging faint-end slopes of the LF (shown here)
and the MF (shown in D13), between the early-type and the global 
galaxy populations at $z<1$ imply that the physical processes that 
switch-off star formation have to be more efficient for massive/luminous galaxies.

This effect is shown equally well by the CMR of Figure~\ref{cmra}, in which
the bright end of the RS is already populated at $z\sim1$, with only a
modest increase in the number of luminous/massive red galaxies at
later epochs. Models of galaxy evolution that can reproduce these trends
can be roughly divided between merger-dominated models, where dry mergers
between galaxies already on the RS contribute significantly to the mass
growth of the brightest and most massive early-type galaxies 
\citep[e.g.,][]{Hop08,SBS12}, and 
quenching-dominated models, where a quenching phase lasting a few Gyr can
steadily transform massive star-forming galaxies into red passive ones, without
the need for significant merging activity.

We have shown in Figure~\ref{LFmp} that a quenching model can very well reproduce
the observed evolution in the RS galaxies number density, as traced by the LF
$\phi^{*}$ parameter, and in Figure~\ref{csm} that the luminous blue galaxies we find 
in the CMR across all explored redshifts ($0.5<z<1$) are in fact massive
objects that provide the natural progenitors for bright red-sequence
galaxies.  These observations suggest that the formation of
new massive quiescent galaxies through merging at $z<1$ is not required, 
as the moderate increase in number density of this population can be explained 
by a few sufficiently massive objects migrating from the blue cloud. 
At the same time, the observed evolution of the LF shows that there must be a
larger increase of the population of quiescent galaxies with
$M_{\star}<10^{11}M_{\odot}$.  
This scenario is also consistent with the low observed rate of dry 
(dissipationless) mergers for galaxies with $M_{\star}>10^{11}M_{\odot}$, with
$\sim$0.5-1 dry merger events per galaxy estimated between $z\sim$1 and today 
\citep{vD05,Bel06,LSLF12}.  

Overall, the build-up of the RS can then be interpreted through a 
continuous supply of objects from the blue cloud through an efficient 
shut down of their star formation over a short time scale 
($\sim$1.5~Gyr, see Section~\ref{sp}) for very massive objects, while at
smaller masses the SF quenching has to be less efficient to explain the time
delay in the population of the fainter part of the RS.
Minor (gas-rich) mergers between these objects can very plausibly play a role
in this process, to explain the observed differences in the internal 
structure of pre-quenched and post-quenched galaxies \citep[e.g.,][]{Bel12}, 
and the size evolution with time of the RS galaxies 
\citep{Dad05,Tru06,Tof07,vDFK08,Sar09,LSLF12,HMS13}.

Additional support for this evolutionary scenario comes from our observation
of a mild increase in the scatter of the RS with redshift (Figure~\ref{uvsca}). 
If there is a regular supply of galaxies migrating from the blue cloud and 
this process was more active at earlier epochs, one naturally expects the RS to appear
broader at higher redshifts.  
Due to a passive fading of the stellar populations which asymptotically
drives the galaxy colours to a typical value (for given metallicity), the 
scatter of the RS decreases and the high mass end of the RS naturally shrinks
as a function of time, while its low-luminosity tail will be extended 
through a delay of star formation in low-mass galaxies preferentially located
in low-density environments \citep{Tan05}.
Our results show that over the range probed by
VIPERS $1>z>0.5$, the intrinsic scatter of the CMR  
decreases with time by $\Delta(U-V)\sim$0.06 mag 
at the bright end ($M_V\lesssim-21$, i.e. roughly $M_V^{*}$ at
$z\sim0.95$), while remaining constant for $M_V\gtrsim-21$.
This is consistent with the luminous end of the CMR being 
build-up over the observed interval (at $z\sim0.45$ the scatter is 
independent of the luminosity), with a small number of additional 
massive members migrating from the blue cloud that do not significantly
alter the average asymptotic red colours. Conversely, at fainter magnitudes
there is an ongoing supply of fresh members over the probed redshift,
with the formation of the faint end being delayed due to the extended star 
formation histories of lower-mass galaxies, in agreement with a 
downsizing formation picture. Our scenario gets independent support 
by the significant buildup of the lower luminosity end of the
red sequence in clusters at $0.4<z<0.8$, whereas the bright 
end is consistent with passive evolution \citep{RvLP09}.

Finally, our interpretation is reinforced by the observation that the 
CMR scatter measured here for the general population is broader than that
measured in cluster environments (Section~\ref{sca}) and the
detection of recent SF activity in the spectra of  
early-type galaxies at high-redshift $z\sim1$ (see Figure~\ref{spec}).
These results are consistent with a scenario in which 
a number of blue low-mass galaxies are pre-processed in denser
environments \citep[groups, e.g.,][]{ZM98,KSNOB01},
before moving to the RS, strongly suggesting that the quenching
mechanism in this case might be related to the environment.

The favoured scenario emerging from our data is somewhat at variance 
with the picture proposed in \cite{Fab07} and \cite{Bel04b},  
in which dry mergers play an important role within a complex formation 
scenario for massive red galaxies that includes a mix of quenching and 
merger processes. The latest incarnation of this model is discussed in 
\cite{SBS12}. Here, the effects of mergers on the evolution of
early-type galaxies are modelled using merger trees in a hierarchical galaxy 
formation framework. These authors distinguish between three models for the 
evolution of a galaxy that has undergone a major merger at high redshift and 
a different evolutionary path at $z<1$: {\em Model~I.} SF is shut-off at 
$z\sim1$ and no dry mergers take place afterwards. In this model the MF does 
not evolve because the number density of RS galaxies does not increase because 
of the lack of wet mergers, and the mix between existing RS galaxies does note
change because of the lack of dry mergers. {\em Model~II.} SF is quenched at 
$z\sim1$ and galaxies undergo on average 0.7 major dry mergers after $z=1$. 
Galaxies that have not yet experienced a major merger at $z>1$
fade passively up to the present. {\em Model~III.} SF continues until galaxies 
undergo a wet major merger. This model is the one closest to the full
semi-analytical model prescriptions and also accounts for the effects of
progenitor bias. However, the model overpredicts the local MF at high masses
($M_{\star}>10^{11.2}M_{\odot}$).  Contrary to the observational results
discussed in \cite{SBS12}, our data show a significant luminosity 
evolution for massive RS galaxies between $z=0.9$ and today, both in
the CMR and in the LF, with $\Delta (U-V)=0.33\pm0.16$ mag and 
$\Delta M_B^{*}=1.07\pm0.09$ mag, respectively.  Both values are in good 
agreement with the predictions of {\em Model~I.} ($\Delta (U-V)\sim0.32$ mag 
and $\Delta M_B\sim1.15$ mag), providing further evidence against a dominant 
role of dry mergers for the build-up of the RS. Of course a very sharp and
highly syncronized cutoff of the SF activity as the one used in this model
cannot be considered very realistic, and if we assume to spread the cutoff
epoch over a finite amount of time we obtain an evolutionary model very similar
to the quenching one discussed in Section ~\ref{LFm}, that is also capable
of reproducing the number density evolution of RS galaxies, as discussed
above, without significantly altering the predictions on the photometric 
evolution of the individual galaxies, and therefore maintainining the agreement
with our observations on the evolution of $M_B^{*}$ and the $(U-V)$ colour
for the RS galaxies. Our results are in agreement with the findings by
\citep{Bro07}, who advocate that $\sim$80\% of the stellar mass of 
very luminous red galaxies ($4L^*$) is already in place at $z=0.7$
and that dry mergers play a minor role in their evolution since $z=0.9$.

This scenario is also consistent with the phenomenological model by
\cite{Pen10,Pen12}, which supports a downsizing effect in the observed 
galaxy properties \citep{GPB96,GS96,CSHC96}. In such a picture, 
a 'mass-quenching' process works on galaxies 
across all mass scales. Such a process is independent of the environment, 
but is proportional to the SFR of each galaxy, which in turn is proportional
to the galaxy stellar mass. A second quenching process, an 'environmental 
quenching', is also part of the model, and is supposed to become effective 
at later epochs, affecting preferentially lower-mass galaxies (essentially 
those not yet affected by the mass-quenching). With the present analysis
we cannot test the impact of the 'environmental quenching' on the
evolution of the RS, but we have clearly shown how a mass-quenching 
scenario describes with good accuracy many aspects of the evolution of the
massive part of the RS. 

Finally, while we provided here a consistent phenomenological scenario
that agrees with the VIPERS observations, the precise physical processes 
that can cause the quenching of star formation are far from being understood. 
They include essentially two possible channels: (1) the
effects of supernova feedback (effective at $M_{h}<5\times10^{11}M_{\odot}$),
virial shock heating \citep{DB2006,Cat06},
or (2) the solution favoured by semi-analytic models, i.e. ``radio-mode"
AGNs \citep{Gra04,Bow06,Cro06,DLJB07}.  Models with a high formation redshift of early-type 
galaxies through gas-rich mergers that experience a phase of quasar activity
\citep[e.g.,][]{Cro06,Bow06,Hop08}
can explain the colour evolution or the MF, but fail to reproduce their 
combined properties for a given redshift \citep{Str09,SBS12,DLB12}.


\section{Summary}\label{sum}

Using a unique sample of approximately 45,000 galaxies with robust spectroscopic 
redshift measurements drawn from the VIPERS PDR-1, we have constructed the
colour-magnitude relation (CMR), the colour-stellar mass relation, and the 
luminosity function (LF) for the total and for the red quiescent galaxy population 
across the redshift range $0.4<z<1.3$. The combination of high-quality 
multi-wavelength data and of a large sample allows us to explore for the first time 
with high accuracy at these redshifts the contribution of the different galaxy populations
to the bivariate distribution of galaxy colours and luminosities.
Our main results can be summarised as follows:

\begin{itemize}

\item The CMR and LF of VIPERS are well populated from $z=1.3$ down to $z=0.4$ 
with a representative number of galaxies of all galaxy types that allows us
to precisely trace the photometric evolution of galaxies over a cosmic period 
of $\sim$4 Gyr. We find that massive red evolved galaxies are already in place
at $z\sim1$, in agreement with previous studies, but the large volume covered by
VIPERS at $z>0.6$, and the resulting unprecedented coverage of the bright end of
the LF, allows us to detect also a significant population of massive blue galaxies 
at high redshift ($0.8\lesssim z\lesssim1.3$). These objects can be considered as 
the natural progenitors of the late massive additions to the RS.
\item The colour and luminosity evolution of the bright part of the RS 
are in agreement with the predictions for a purely passively evolving old
stellar population. We measure a colour evolution of $\Delta (U-V)=0.33\pm0.16$ 
mag and  an evolution of $\Delta M_B^{*}=1.07\pm0.09$ mag since $z=1$.
These results are consistent with the no-merger model predictions for 
massive galaxies by \cite{SBS12}, and show an evolution of the RS
galaxies significantly larger than most merger-dominated evolution models
would predict.
\item The observed evolution of the RS favours stellar population models with
exponentially declining SFHs with $0.3\le\tau\le0.8$ and a formation 
epoch at $2.5<z<3.5$. Our observations rule out SFHs with $\tau>0.8$ 
and require a rapid build-up of the RS over a very short time scale of 
only $\sim$1.5 Gyr; clearly, this has to happen at epochs earlier than
our redshift limit of $z=1.3$.
The measured RS evolution gives strong constraints on the predictions of
stellar population models and suggests weak star formation episodes over
the redshift range $0.8\lesssim z\lesssim1.2$. This claim is supported by 
residual star formation detected in the stacked spectra of early-type galaxies 
at high-redshift ($0.9<z<1.0$) which involves only a few percent of 
their total stellar mass. In contrast, stacked spectra of early-type 
galaxies at low-redshift ($0.5<z<0.6$) show the signature of a purely quiescent 
stellar population with no signs of recent star formation activity.
Moreover, the evolution of dusty red galaxies is mild up to $z=1$ and 
the influence of dust obscuration on the internal properties of quiescent
galaxies becomes important at $z>1.3$.
\item The intrinsic scatter of the RS for the galaxy sample based on the 
SED-type classification is increasing from $z=0.4$ to $z=1.1$. The total intrinsic
scatter of the CMR is $\sigma(U-V)\sim0.11$, which is a factor of two larger 
than the scatter of the CMR found in galaxy clusters at the same redshifts. Our
observed intrinsic scatter in the field is not in agreement with theoretical
model predictions \citep{Men08}, which suggest a scatter that is three 
times as large as observed.
\item The significant luminosity and number density evolution measured in 
the LF supports a formation scenario where massive early-type galaxies 
($M_{\star}>10^{11}M_{\odot}$) have been assembled through wet mergers at an 
early epoch in the universe, followed by a powerful shut-down mechanism 
(such as quenching) of their star formation activity at $z\sim1$ over a short 
time-scale. Afterwards the luminosity of these systems quickly fades, and once 
on the RS, they evolve in a purely passive way with their stellar populations 
continuously becoming fainter and redder.
\item Our results imply that gas-poor (dry) major-mergers of massive galaxies
are not the dominant factor in the build-up the bright, massive end of the RS since
$z\sim1$ and they also do not contribute to the number density of massive
quiescent galaxies. Massive blue star-forming galaxies that exist at $z\sim1$
can be transformed into passive quiescent RS galaxies through some quenching
mechanism. Mergers however are still likely to contribute to the evolution
of the internal structure and the size of galaxies on the RS.
\end{itemize}


\begin{acknowledgements}
  We would like to thank the anonymous referee for a constructive review of 
  this manuscript. We acknowledge the crucial contribution of the ESO staff for
  the management of service observations. In particular, we are deeply grateful to 
  M. Hilker for his constant help and support of this programme. Italian participation 
  to VIPERS has been funded by INAF through PRIN 2008 and 2010 programs. LG and
  BRG acknowledge support of the European Research Council through the
  Darklight ERC Advanced Research Grant (\# 291521). OLF acknowledges support
  of the European Research Council through the EARLY ERC Advanced Research
  Grant (\# 268107). Polish participants have been supported by the Polish
  Ministry of Science (grant N N203 51 29 38), the Polish-Swiss Astro Project
  (co-financed by a grant from Switzerland, through the Swiss Contribution to
  the enlarged European Union), the European Associated Laboratory Astrophysics
  Poland-France HECOLS and a Japan Society for the Promotion of Science (JSPS)
  Postdoctoral Fellowship for Foreign Researchers (P11802). GDL acknowledges
  financial support from the European Research Council under the European
  Community's Seventh Framework Programme (FP7/2007-2013)/ERC grant agreement
  n. 202781. WJP and RT acknowledge financial support from the European Research
  Council under the European Community's Seventh Framework Programme
  (FP7/2007-2013)/ERC grant agreement n. 202686. WJP is also grateful for
  support from the UK Science and Technology Facilities Council through the
  grant ST/I001204/1. EB, FM and LM acknowledge the support from grants
  ASI-INAF I/023/12/0 and PRIN MIUR 2010-2011. LM also acknowledges 
  financial support from PRIN INAF 2012. YM acknowledges support from
  CNRS/INSU (Institut National des Sciences de l'Univers) and the
  Programme National Galaxies et Cosmologie (PNCG). CM is grateful for
  support from specific project funding of the {\it Institut Universitaire
  de France} and the LABEX OCEVU.
  This research uses data from the VIMOS VLT Deep Survey, obtained from the 
  VVDS database operated by Cesam, Laboratoire d'Astrophysique de Marseille, 
  France.
\end{acknowledgements}


\begin{appendix} 

\section{The Johnson-Cousins $UBVRI$ System}\label{uband}

%
%
\begin{figure}
 \centering
\includegraphics[width=0.37\textwidth, angle=-90]{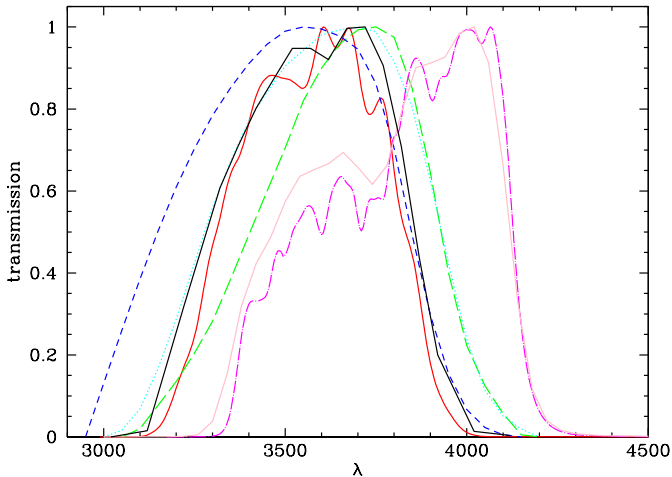}
 \caption{Filter transmission curve for different popular $U$-band filters in the
 literature. The $U$-Johnson (blue dashed), used in VIPERS, is compared to the
 $U_{\rm Bessel}$ VVDS filter (red solid), $U3$ Buser filter (green),
 $U_{\rm Bessel 1990}$ (cyan dotted), $u$ SDSS (black), $u^{*}$ CFHT MegaCam
 (pink), and the $u^{\star}$ CFHT MegaCam SAGEM filter (magenta dot-dashed line).}
 \label{Utrans}
\end{figure}

One of the most common and frequently used standard broad-band photometric
systems is the photoelectric $UBV$ system. Usually, the Johnson-Cousins
$UBVRI$ refers to the combined Johnson-Cousins $UBV$ system \citep{JM1953}
and its red optical extension of the Cousins $RI$ system \citep{C1976}. 
In this system the $V$-band represents an approximate measure of the visual
photographic magnitude, whereas the $B$-band was defined to give a measure for the
uncorrected photographic magnitude. In addition, the $U$-band probes the
interesting window between the atmospheric cutoff and the $B$-band. However,
the combination of the $UBV$ system of having a cutoff defined by the atmosphere
at low wavelengths (plus the original 1P21 glass optics) and a limit by the detector
at long wavelengths ($\lambda\lambda\approx 6320\approx V_{{\rm Rcut}}$) implicated that the $U$-band filter is
sensitive to the atmospheric extinction. The original $B$ and $V$ bands of the $UBV$
system could be well reproduced with current more sentitive (redder) CCD detectors
\cite[e.g.,][for the $U$-band]{B1978,B1986} \cite[][for $UBVRI$]{B1990}. However, the
characteristics of the $U$-band were more difficult to realize with current detector
technologies, mainly because of short wavelength cutoff due to the atmosphere and
the the impact of temperature dependencies \citep{BCP98}.

In the photometric Johnson system the A0V star $\alpha$ Lyr (Vega) is defined to
have $V$=0.03 mag, whereas all other colours of Vega are equal to zero. The absolute
flux of Vega can be calibrated using empirical relations 
\citep[e.g.,][]{B1979,B1990}, or some calibration standard stars \citep{Land1992}.

\begin{table}[t]
\caption{Basic characteristics of different $UBV$ filters. 
Columns show the name of the filter, the central wavelength $\lambda_c$, 
the effective wavelength $\lambda_o$, the Width-at-Half-Maximum (WHM),
and the effective width $W_o$.}
\label{filt}
\centering                         
\begin{tabular}{c c c c c c}   
\hline\hline
Filter & $\lambda_c$ & $\lambda_o$ & $\lambda_p$ & WHM & $W_o$  \\
     & \AA & \AA  & \AA & \AA  & \AA \\
\hline
$U$ photoel  & 3499 & 3502 & 3550 & 699 & 681 \\  
$U3$ Buser   & 3666 & 3652 & 3754 & 525 & 543 \\  
$B3$ Buser   & 4368 & 4417 & 4150 & 958 & 974 \\  
$V$ Buser    & 5426 & 5505 & 5285 & 827 & 870 \\  
\hline                        
\end{tabular}
\tablefoot{Data taken from Asiago database:
http://ulisse.pd.astro.it/Astro/ADPS}
\end{table}

For the present work, we have adopted the original 1953 $U_J$-Johnson filter
($\lambda_c=3499$, WHM=699, hereafter $U$-band) that has been reconstructed in
the USA \citep{JM1953}. The original filter name is referred as
Corning 9863 and was initially used in photoelectric observations. 
For an unreddened A0V star with $V=0.00$, the $U$-Johnson filter gives a total
flux of $3.98\times 10^{-9}$~erg cm$^{-2}$ sec$^{-1}$ \AA$^{-1}$ \citep{L1982}. Our
choice is primarily driven by the high filter response in the blue wavelength
range compared to other $U$-band filters used in the literature. Because of its
high sensitivity in the blue, the $U$-Johnson filter allow us to directly probe
the luminous, hot and blue OB stars and SF associations in the stellar content
of galaxies and therefore acts as a proxy for SF, although it is affected by 
dust extinction. However, using a combination of GALEX $NUV$ and $FUV$ IR
colours, we demonstrate in Section~\ref{nuv} that we are able to perform a
robust separation into red quiescent, blue star-forming galaxies as well as
dust-obscured red galaxies with or without SF. Similar approaches adopting a
pseudo-continuum $U_{280}$ filter to split quiescent from star-forming galaxies
have been used, for example, in moderate redshift clusters \citep{WAB09}
and for field galaxies \citep{Nic11}.

Figure~\ref{Utrans} shows the filter transmission curve for the $U$-Johnson filter
compared to other frequently used $U$-band filters in the literature. The
$U$-Johnson (blue dashed) is compared to the $U_{\rm JKC}$ Bessel filter (red solid),
$U3$ Buser filter (green), $U_{\rm Bessel\,1990}$ (cyan dotted), $u$ SDSS (black),
$u^{*}$CFHT MegaCam (pink), and the $u^{\star}$ CFHT MegaCam SAGEM filter
(magenta  dot-dashed line), which is used in ALF, whereas for the
the SED modelling we adopt $U$-Johnson filter.
Note that the Johnson filter has a quite different efficiency curve from the
standard $U_{\rm JKC}$-Johnson-Kron-Cousins filter in the
VVDS, or the $U3$ Buser filter as adopted in the PEGASE filter library
\cite[see Table~1 of][]{BK1978} or in the \cite{BC03}
models (record 12). The $U$-Johnson has a higher blue sensitivity than any other
$U$-band filter and allows to obtain redder colours for red galaxies with a more
prominent separation between blue and red galaxies with a minimum at 
$(U-B)\sim 1.2$ (AB). Moreover, the whole filter remains below 4000\,\AA\ and
therefore is a good proxy for the 4000\,\AA\ break. The $U3$ filter is extended at
larger wavelengths and therefore produces bluer colours compared to the
$U$-Johnson filter and a minimum at  $(U3-B)\sim 1.0$ (AB).

For the $B_{JKC}$ filter we adopt the $B3$ Buser filter \citep{B1978} as 
commonly used in PEGASE ($B3_{BK78}$) and the \cite{BC03} models (record 14). As 
the $V_{JKC}$ filter we define the $V$ Buser (corresponding to $V_{BK78}$ in
PEGASE and record 15 in the \cite{BC03} models). Both the $B3$ and $V$ 
Buser filters are also used in the Millennium simulation by \cite{DLJB07}. 
Table~\ref{filt} compares the main filter characteristics used in this work 
($U,B3,V$) to the $U3$ filter.  

We adopt the following transformation from the $U_J$-Johnson filter passband
to the rest-frame $U_{JKC}$ system (in AB):
\begin{equation}
U_{JKC} = 1.00(\pm0.02)\times\ U_J-0.14(\pm0.02)\times(U_J-B_J)+0.12(\pm0.01)
\end{equation}
The transformation from AB to Vega system was performed using 
$M_{JKC}{\rm(AB)}=M_{JKC}{\rm(Vega)}+c_{X}{\rm(AB)}$, where the individual
colour terms $c_{X}{\rm(AB)}$ for each filter were derived
through the SED fitting procedure.

Among the literature the interpretation and application of the 
$UBV$-Johnson-Morgan-Cousins system is often inhomogeneous and ambiguous.
For example, \cite{CCG10} used the $U3$ \citep{B1978} filter, whereas
the $B$ and $V$-band are not the corresponding Buser filters but the 
$B2$ and $V$ filters by \cite{AS69}.
More popular are the usage of the $UBV_{JKC}$ Johnson-Cousins definitions,
like for the VVDS \citep{Fra07} or in the zCOSMOS survey 
\citep{Cuc10}. It is beyond the scope of the current investigation to
reproduce the exact filter definitions used among works in the literature. We emphasize
that that for comparisons with literature data one should always be precise 
and clearly describe which photometry and filter transmission curves are adopted.

\section{Completeness Test}\label{compl} 

In Figure~\ref{CMRev} we observe a change of the evolution of the RS intercept
from $z=0.9$ to $z=1.3$. This change in the number of red galaxies could be
either due to a real evolution or to sample incompleteness. 
To test the completeness of red galaxies in the highest redshift bin 
$1.0<z<1.3$, we construct different samples of red galaxies in the lower 
redshift bin $0.9<z<1.0$ (hereafter simulated samples) to verify whether the
observed properties of the sample at $1.0<z<1.3$ (hereafter real sample) are
consistent with the properties of their counterparts at $0.9<z<1.0$. 
Red galaxies in VIPERS are defined as those galaxies classified by SED type 
class 1 (see Section~\ref{type}). We assume that the average observed $(U-V)$ 
rest-frame distribution is the same for red galaxies at all redshifts.
Because the reddest galaxies have the faintest ultra-violet magnitudes, 
a possible incompleteness bias would be apparent in the observed $i'$-band
magnitude distribution with the reddest galaxies being absent.

We have computed the observed $(U-V)$ rest-frame distribution of 
randomly selected red samples in the redshift bin 
$0.9<z<1.0$, which were extracted from the observed $(U-V)$ rest-frame 
distribution of all red galaxies within the same redshift interval. The real 
and randomly selected red galaxy samples at $0.9<z<1.0$ share the same
properties and the randomly selected (hence simulated) samples are always
a sub-set of the total red galaxy sample. These simulated samples 
have the same number of galaxies as the sample at $1.0<z<1.3$ and 
therefore should mimic the properties of the observed red galaxy sample
at the highest redshift bin, assuming the latter is complete. 
We take a random set of two simulated samples at $0.9<z<1.0$, referred to
simulated 1 and simulated 2, to understand their variance and possible spread
in properties. We assume the redshift bin $0.9<z<1.0$ to be complete for all
types of (red, green and blue) galaxies and we are mainly interested in possible
incompleteness effects in the RS evolution at $z>1$. For each galaxy
in the simulated samples we derive the observed $i'$-band (AB) magnitudes.

For the real galaxy sample, we took the observed $(U-V)$ rest-frame 
distribution of all red galaxies in the high-redshift bin $1.0<z<1.3$ 
and computed for each galaxy separately the observed $i'$-band 
magnitudes. Finally, we corrected for the redshift evolution of each 
object as $\Delta(M_{corr})=\Delta(M_{i'})/0.3\times(z_{\rm spec}-1)$,
where $M_{i'}$ is the evolution correction from $z=1.3$ to $z=0.9$
derived from the LF of the VVDS \citep{ITZ05}  
and $z_{\rm spec}$ is the spectroscopic redshift of each single galaxy.

\begin{figure}
 \centering
 \includegraphics[width=0.5\textwidth]{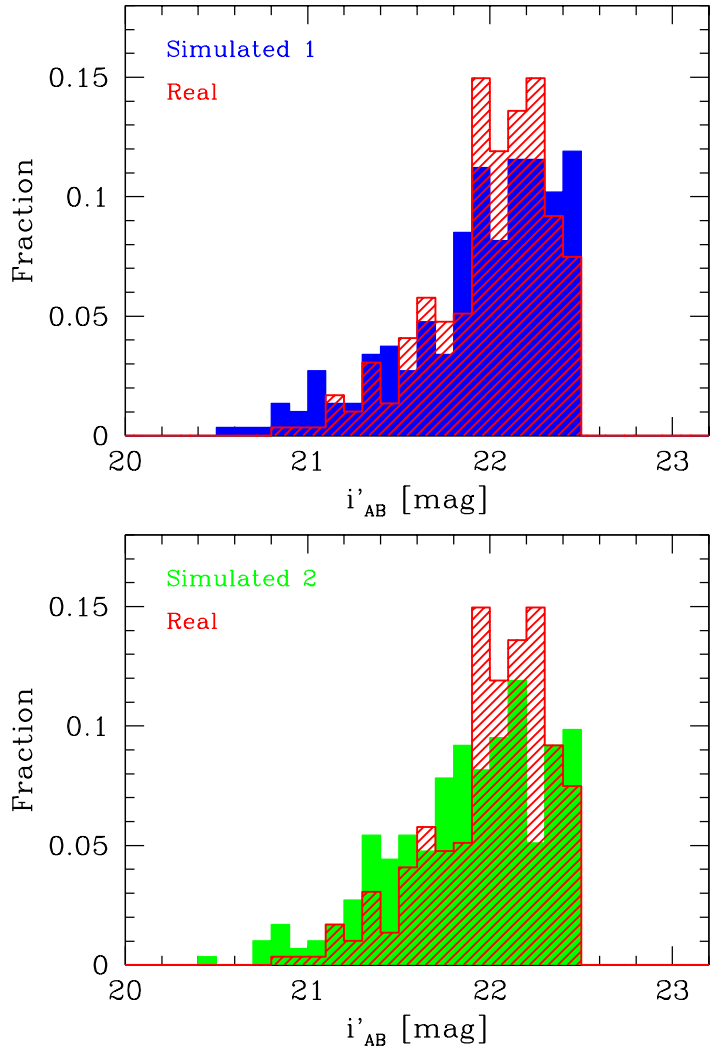}
 \caption{Completeness test for red galaxies in the VIPERS PDR-1.
 The red histogram displays the observed real red galaxies at $1.0<z<1.3$
 transformed to $0.9<z<1.0$, whereas the simulated samples are two 
 representations of red galaxy samples at $0.9<z<1.0$. The properties of 
 the real sample are consistent with the properties of their 
 simulated counterparts.}
 \label{icompl}
\end{figure}

A comparison of the real and simulated red galaxy samples is shown 
in Figure~\ref{icompl}. Two representative simulated
samples of red galaxies (simulated 1 and simulated 2) are shown in blue and
green, respectively. The red histogram shows the real red sample at 
$1.0<z<1.3$, which was transformed to redshift $0.9<z<1.0$. Table~\ref{stat} 
compares the median, first and third quartile, and the 1~$\sigma$ Gaussian
values of the distribution for the real and randomly selected simulated samples.
The histograms display similar shapes and show consistent statistics. We 
therefore conclude that our red galaxy sample at $z>1$ does not show any 
significant incompleteness of bright galaxies and that our sample is 
also highly complete at fainter magnitudes.

\begin{table}[t]
\caption{Statistical properties of real and simulated red galaxy samples.}
\label{stat}
\centering                         
\begin{tabular}{c c c c}   
\hline\hline
Quantity & Real & Simulated 1 (blue) & Simulated 2 (green) \\
\hline
median      &   22.07   &   22.05    &    21.95 \\
quartile~1  &   21.84   &   21.74    &    21.62 \\
quartile~3  &   22.24   &   22.27    &    22.19 \\
$\sigma$    &   0.33    &   0.42    &    0.42 \\
\hline                        
\end{tabular}
\end{table}

%
%

\section{Cosmic Variance}\label{cvar} 

Statistical measurements based on number counts such as the luminosity
function or mass function are subjected to field-to-field variations of the
number density that originate from the clustering of a particular galaxy 
population and from variations imprinted by the scale of the probed 
survey volume.

For a probability distribution function $P_N(V)$ that denotes the 
probability of counting $N$ objects within a volume $V$, the relative
cosmic variance is defined as
\begin{equation}
\sigma^{2}_{\rm cv}=\frac{\langle\ N^{2}\rangle-\langle\ N\rangle^{2}}{\langle\ N\rangle^{2}}-
\frac{1}{\langle\ N\rangle},
\end{equation}
where $\langle\ N\rangle$ and $\langle\ N^{2}\rangle$ are the mean and variance 
of the galaxy number counts \citep{SLF04}.

\begin{figure*}
 \centering
 \includegraphics[width=0.8\textwidth]{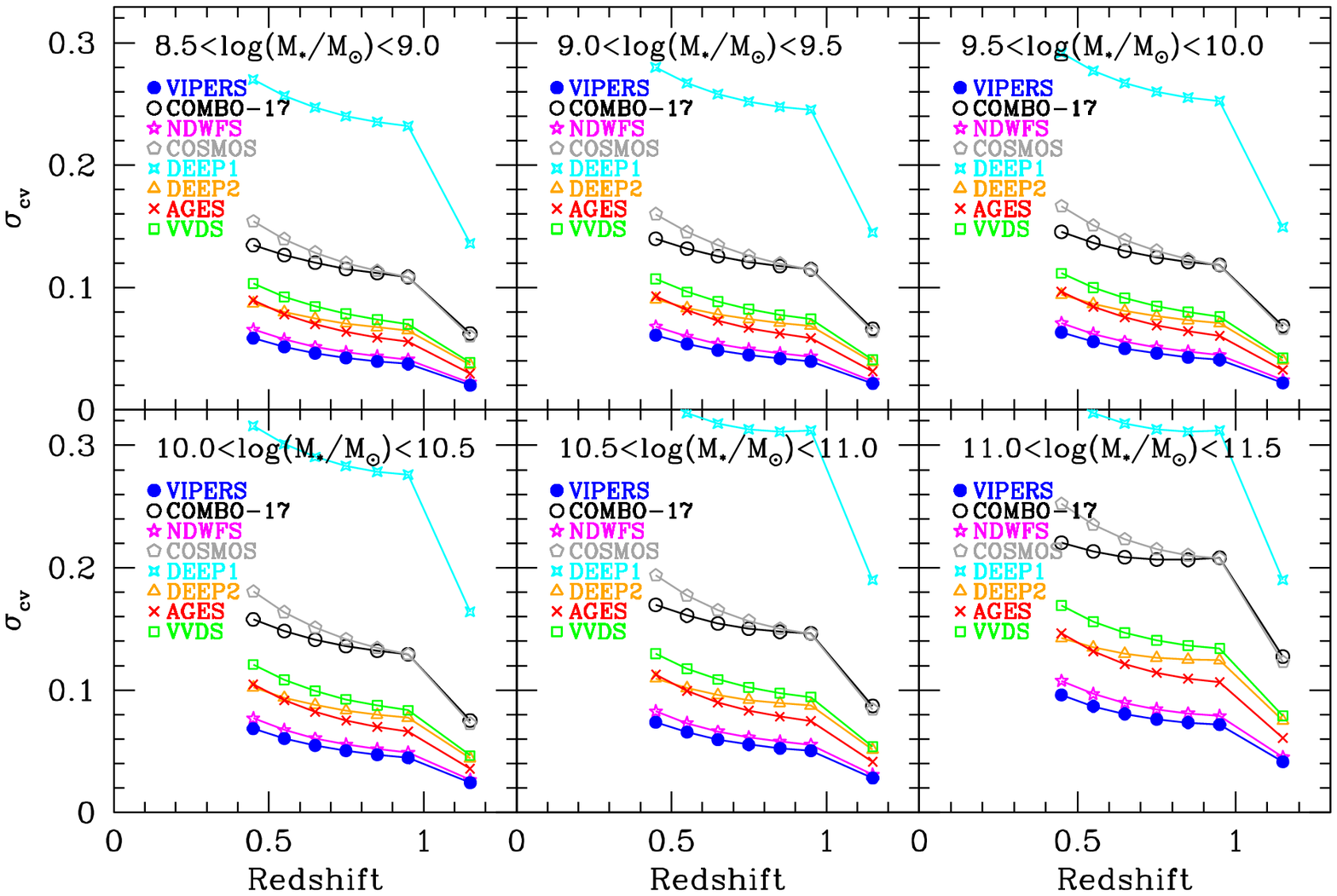}
 \caption{Cosmic variance for the VIPERS PDR-1 sample and other surveys
 from the literature. Symbols denote data from 
 AGES \citep[][red crosses]{CEK12}, COMBO-17 \citep[][black circles]{Bel04b},
 COSMOS \citep[][grey polygons]{SAAB07}, DEEP1 \citep[][cyan stars]{Im02},
 DEEP2 \citep[][orange triangles]{Fab07}, NDWFS \citep[][magenta stars]{Bro07},
 and VVDS \citep[][green squares]{VVDS05,VVDS13}.}
 \label{cv}
\end{figure*}

To test the impact of cosmic variance on our results, we have 
computed the uncertainty of cosmic variance for the VIPERS survey
using the public code {\tt getcv} \citep{MSNR11}.  
Figure~\ref{cv} shows the relative cosmic variance uncertainty in
VIPERS (blue filled circles) divided into different mass ranges probed by 
the survey. For reference purposes, we also show the results of several 
other surveys taken from the literature.

For RS galaxies between $0.4<z<1.3$, the uncertainties arising from cosmic 
variance vary in the range $0.04<\log(M_{\star}/M_{\odot})<0.07$, with 
a median of $\langle\log(M_{\star}/M_{\odot})\rangle=0.05$.
Note that the effective area of VIPERS is 10.32~deg$^2$, which is about 
4~deg$^2$ larger than any other survey at intermediate redshifts (e.g., NDWFS, AGES).
Figure~\ref{cv} shows that independent of the mass probed, the cosmic variance
effects on VIPERS are a factor of two lower than AGES and 20\% lower than NDWFS.
Compared to all the other surveys, VIPERS is a factor of 4 or more less 
affected to cosmic variance effects. In particular, compared to DEEP2 and
COMBO-17, the VIPERS data offers a huge improvement. Recent surveys like 
NDWFS or AGES cannot compete either with VIPERS. We emphasize that AGES has
an average sampling rate of 20\% in $I$ and $K$-bands \citep{CEK12}, which is 
a factor of two lower than the median sampling rate of VIPERS. We therefore
conclude that the impact of cosmic variance has a negligible effect on our
results.

%
%

\end{appendix}

\end{document}